%% file: main.tex
\documentclass[sigplan,twocolumn,screen]{acmart}

\renewcommand\footnotetextcopyrightpermission[1]{}
\acmConference[EuroSys '26] {The 21st European Conference on Computer Systems}{April 27--30, 2026}{Edinburgh, UK.}
\acmBooktitle{The 21st European Conference on Computer Systems (EuroSys '26), April 27--30, 2026, Edinburgh, UK}

\acmSubmissionID{96}

\input{misc/config.tex}
\input{misc/macro.tex}
\input{misc/authors}

\renewcommand{\shortauthors}{Zhixin Zhao et al.}

\begin{document}

\title[PARD: Enhancing Goodput for Inference Pipeline via Proactive Request Dropping]{PARD: Enhancing Goodput for Inference Pipeline via \underline{P}ro\underline{A}ctive \underline{R}equest \underline{D}ropping}


\input{content/_revision}
\input{content/0_abstract}

\maketitle

\input{content/1_introduction}

\input{content/2_background}
\input{content/3_motivation}

\input{content/4_design}

\input{content/5_evaluation}
\input{content/6_related}

\input{content/7_conclusion}

\bibliographystyle{plain}
\bibliography{misc/sample}

\end{document}

%% file: misc/config.tex
\usepackage[]{hyperref} 
\usepackage{bm}
\usepackage{xspace}
\usepackage{pifont}
\usepackage{color}
\usepackage{amsmath}
\usepackage{makecell}
\usepackage{wrapfig}
\usepackage{booktabs}
\usepackage[skip=3pt]{subcaption}
\usepackage{enumitem}
\usepackage{microtype}
\usepackage[normalem]{ulem} 
\usepackage[ruled,vlined]{algorithm2e}
\usepackage{trimspaces}

\newcommand{\narrow}[1][-9pt]{\vspace{#1}}

\makeatletter
\g@addto@macro{\UrlBreaks}{\UrlOrds}
\makeatother

%% file: misc/macro.tex
\newcommand{\sysname}{PARD\xspace}

\newif\ifshowtodos

\usepackage[textsize=small]{todonotes}

\newif\ifshowrevise

\newif\ifshoworigin

\newcommand{\response}[2]{
}

\newcommand{\revise}[2]{%
  \ifshowrevise
\else#2\fi
}

\newcommand\paraspace{}
\providecommand\parab[1]{\addvspace{3pt}\noindent\textbf{#1}}
\providecommand\parae[1]{\paraspace\textbf{\textit{#1}}}

\providecommand\revision[1]{\addvspace{9pt}\noindent\textbf{#1}\medskip}

\newcommand{\ie}{\emph{i.e.,}\xspace}
\newcommand{\eg}{\emph{e.g.,}\xspace}

\newcommand{\secref}[1]{\S\ref{#1}}
\newcommand{\figref}[1]{Figure~\ref{#1}}
\newcommand{\tabref}[1]{Table~\ref{#1}}
\newcommand{\algoref}[1]{Algorithm~\ref{#1}}
\newcommand{\eqnref}[1]{Equation~\ref{#1}}

\newcommand{\page}[1]{Page~\pageref{#1}}

%% file: misc/authors.tex
\author{Zhixin Zhao}
\orcid{0009-0009-7865-4905}
\affiliation{%
  \institution{Tianjin University}
  \city{Tianjin}
  \country{China}
}
\email{zhao612@tju.edu.cn}

\author{Yitao Hu}
\orcid{0009-0004-0458-0900}
\authornote{Corresponding author.}
\affiliation{%
  \institution{Tianjin University}
  \city{Tianjin}
  \country{China}
}
\email{yitao@tju.edu.cn}

\author{Simin Chen}
\orcid{0000-0001-5035-3398}
\affiliation{%
  \institution{University of Texas at Dallas}
  \city{Richardson}
  \country{USA}
}
\email{simin.chen@utdallas.edu}

\author{Mingfang Ji}
\orcid{0009-0007-0691-7994}
\affiliation{%
  \institution{Tianjin University}
  \city{Tianjin}
  \country{China}
}
\email{jimingfang@tju.edu.cn}

\author{Wei Yang}
\orcid{0000-0002-5338-7347}
\affiliation{%
  \institution{University of Texas at Dallas}
  \city{Richardson}
  \country{USA}
}
\email{wei.yang@utdallas.edu}

\author{Yuhao Zhang}
\orcid{0000-0002-0118-8135}
\affiliation{%
  \institution{Tianjin University}
  \city{Tianjin}
  \country{China}
}
\email{yuhaozhang@tju.edu.cn}

\author{Laiping Zhao}
\orcid{0000-0003-1967-2192}
\affiliation{%
  \institution{Tianjin University}
  \city{Tianjin}
  \country{China}
}
\email{laiping@tju.edu.cn}

\author{Wenxin Li}
\orcid{0000-0001-8507-0339}
\affiliation{%
  \institution{Tianjin University}
  \city{Tianjin}
  \country{China}
}
\email{toliwenxin@tju.edu.cn}

\author{Xiulong Liu}
\orcid{0000-0001-8414-1668}
\affiliation{%
  \institution{Tianjin University}
  \city{Tianjin}
  \country{China}
}
\email{xiulong_liu@tju.edu.cn}

\author{Wenyu Qu}
\orcid{0009-0005-3773-4619}
\affiliation{%
  \institution{Tianjin University}
  \city{Tianjin}
  \country{China}
}
\email{wenyu.qu@tju.edu.cn}

\author{Hao Wang}
\orcid{0000-0002-1444-2657}
\affiliation{%
  \institution{Stevens Institute of Technology}
  \city{Hoboken}
  \country{USA}
}
\email{hwang9@stevens.edu}

%% file: content/_revision.tex
\setcounter{page}{1}

%% file: content/0_abstract.tex
\begin{abstract}


Modern deep neural network (DNN) applications integrate multiple DNN models into inference pipelines with stringent latency requirements for customized tasks. To mitigate extensive request timeouts caused by accumulation, systems for inference pipelines commonly drop a subset of requests so the remaining ones can satisfy latency constraints. Since it is commonly believed that request dropping adversely affects goodput, existing systems only drop requests when they have to, which we call \textit{reactive} dropping. However, this reactive policy can \textit{not} maintain high goodput, as it neither makes timely dropping decisions nor identifies the proper set of requests to drop, leading to issues of dropping requests too late or dropping the wrong set of requests.

We propose that the inference system should \textit{proactively} drop certain requests in advance to enhance the goodput across the entire workload. To achieve this, we design an inference system \sysname. It enhances goodput with timely and precise dropping decisions by integrating a \revise{bi-directional}{proactive} dropping method that decides when to drop requests using runtime information of the inference pipeline, and an adaptive request priority mechanism that selects which specific requests to drop based on remaining latency budgets and workload intensity. Evaluation on a cluster of 64 GPUs over real-world workloads shows that \revise{\sysname achieves $21\%$--$176\%$ higher goodput than the state of the art while reducing the drop rate and wasted computation resources by $1.6\times$--$16.7\times$ and $1.5\times$--$62\times$ respectively.}{\sysname achieves $16\%$--$176\%$ higher goodput than the state of the art while reducing the drop rate and wasted computation resources by $1.6\times$--$17\times$ and $1.5\times$--$62\times$ respectively.}

\end{abstract}

%% file: content/1_introduction.tex
\newpage

\section{Introduction}
\label{sec:intro}

\begin{figure}[t]
    \centering
    \includegraphics[width=0.48\textwidth]{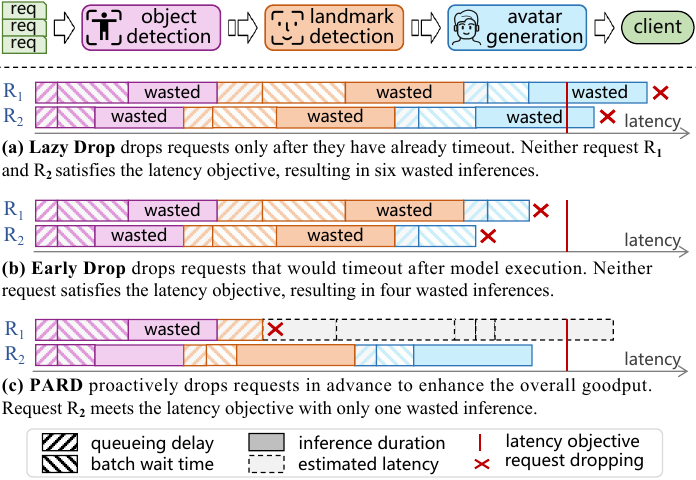}
    \caption{Request latency composition under various dropping policies in a three-model inference pipeline.}\narrow{}
    \label{fig:intro_dropping}
    \Description{}
\end{figure}



\response{B/D/E}{Rename the paper title and remove "bi-directional".}

Increasing task complexity and proliferation of open-source models have made DNN inference pipelines crucial and cost-efficient for emerging real-time applications~\cite{shen2019nexus, hu2021scrooge, razavi2022fa2, ghafouri2023ipa, crankshaw2020inferline, hsieh2018focus, xiang2019pipelined}. These pipelines consist of multiple DNN models, where requests pass through each model to complete customized tasks under a strict latency objective. For example, the avatar generation pipeline~\cite{hu2021scrooge, shen2019nexus} in \figref{fig:intro_dropping} transforms live video of a person into a virtual character for real-time interaction. Existing inference systems define goodput~\cite{zhang2019mark, zhang2023shepherd, gunasekaran2022cocktail}, which is the number of requests that can meet latency objective per unit of time, to measure how well the system provides services for latency-sensitive inference pipelines. To guarantee high goodput, existing systems primarily focus on techniques such as resource scaling~\cite{ahmad2024proteus, zhang2019mark, xu2022igniter, zhang2020enabling}, dynamic batching~\cite{shen2019nexus, hu2021scrooge, gao2018low, zhang2023octopus} and GPU scheduling~\cite{xia2023towards, pang2023efficient, yu2021automated}.



\revise{
    Despite these efforts, certain requests still encounter large latency due to unpredictable events such as workload bursts [cite: zhang2019mark, gunasekaran2022cocktail] or machine failure [cite: dean2013tail].
}{%
    Despite these efforts, certain requests inevitably experience large latencies due to unpredictable events such as workload bursts~\cite{zhang2019mark, gunasekaran2022cocktail} or machine failure~\cite{dean2013tail}. In real-time DNN serving (\eg video analytics), requests that fail to meet their latency service-level objectives (SLOs) are useless~\cite{crankshaw2020inferline, kim2023dream, shen2019nexus}.
}
Without careful handling, these requests prolong the latency of subsequent requests due to the queueing effect, eventually leading to latency violations and poor goodput. Therefore, existing systems employ request dropping policies to drop a portion of requests~\cite{shen2019nexus, ghafouri2023ipa, hu2021scrooge, crankshaw2017clipper, gujarati2017swayam}, ensuring the remaining ones meet the latency objective and avoid goodput degradation.

\response{C}{Clarify why dropping timeout requests.}


However, existing dropping policies can \textit{not} maintain high goodput since they share a reactive design. Since it is commonly believed that dropping requests adversely affects goodput, existing systems \textit{drop requests only when they have to be dropped}, \ie when it has already timed out before DNN model execution~\cite{crankshaw2017clipper} (\figref{fig:intro_dropping}a) or will be timed out after its execution~\cite{shen2019nexus} (\figref{fig:intro_dropping}b). These reactive designs fail to make proper dropping decisions for two primary reasons. (1) Drop requests too late. When making decisions at a given module\footnote{Each module serves a specific DNN model of the inference pipeline using the assigned computation resources.} in the pipeline, the reactive policy only considers the latency up to the current module. It ignores the latency budget for subsequent modules, leading to inopportune dropping decisions and wasted computation resources. (2) Drop the wrong set of requests. The reactive policy makes dropping decisions in request's arrival order. It ignores the impact of the difference in requests' remaining latency budget and the workload variation, which cannot determine the proper set of requests to drop for goodput enhancement.

In this paper, we argue that \textbf{inference system should proactively drop certain requests in advance to enhance the goodput for the entire workload}. However, three characteristics of the inference pipeline make this challenging. (1) \textit{Model Cascading.} Requests traverse multiple cascaded models with only end-to-end latency objectives specified. While the latency budget of each model must be considered when making dropping decisions. (2) \textit{Latency Uncertainty.} Batch wait time from the batching process~\cite{shen2019nexus} and queueing cause high uncertainty in request latency. This uncertainty is further amplified by model cascading of the pipeline. (3) \textit{Workload Variation.} Inference workloads exhibit dynamic and bursty characteristics, demanding that dropping policy effectively adapts to varying workload intensity. Building on our argument, we design and implement an inference system called \sysname, which adapts the characteristics and makes proactive dropping decisions to enhance the goodput of the inference pipeline as follows.

\revise{
    First, \sysname designs a novel \revise{bi-directional}{proactive} request dropping method to address \textit{when to drop requests} for each module in the inference pipeline. Existing reactive policies rely solely on unidirectional latency information from preceding modules and overlook the latency budget requirements of subsequent modules, resulting in most dropped requests being concentrated in the last few modules of the pipeline. On the contrary, \sysname proactively estimates the latency of requests in the entire pipeline with bi-directional runtime information from the preceding and subsequent modules, so as to make timely dropping decisions in the early stages of the pipeline, reducing the waste of computation resources (\figref{fig:intro_dropping}c).
}{%
    First, \sysname designs a novel \revise{bi-directional}{proactive} request dropping method to address \textit{when to drop requests} for each module in the pipeline. Existing reactive policies rely solely on latency information from preceding modules and overlook the latency budget requirements of subsequent modules, causing most drops to concentrate in the last few modules of the pipeline. In contrast, \sysname proactively estimates request latency across the entire pipeline with bi-directional runtime information from the preceding and subsequent modules, enabling timely dropping decisions in the early stages of the pipeline, reducing wasted computation (\figref{fig:intro_dropping}c).
}

\revise{
    Second, \sysname leverages a novel adaptive request priority method to address \textit{which specific requests to drop} for each module in the inference pipeline. The existing reactive policies make dropping decisions in requests' arrival order, which cannot reflect the difference in their remaining latency budget. On the contrary, \sysname designs a hybrid priority mechanism that determines the decision-making order across requests based on each request's remaining latency budget and workload intensity. This mechanism ensures a smooth transition between dropping priorities, so as to determine a proper set of requests to drop under dynamic workloads, which avoids unnecessary request dropping.

}{
    Second, \sysname introduces an adaptive request priority method to decide \textit{which specific requests to drop} in each module. Existing reactive policies drop requests strictly by arrival order, ignoring the difference in their remaining latency budget. In contrast, \sysname designs a hybrid priority mechanism that determines the decision-making order across requests based on each request's remaining latency budget and workload intensity. This mechanism ensures a smooth transition between dropping priorities, so as to determine a proper set of requests to drop under dynamic workloads, which avoids unnecessary request dropping.
}

\response{none}{Shorten the introduction to reserve space for \secref{sec:eval}-\secref{sec:discussion}.}

When enhancing the goodput for the inference pipeline, \sysname makes the following contributions:

\begin{itemize}[leftmargin=12pt,labelsep=6pt,topsep=0pt,itemsep=0pt]
    \item \sysname highlights the critical role of request dropping in enhancing goodput and reveals why existing reactive design fails to maintain high goodput for inference pipeline through a systematic analysis.
    \item \revise{\sysname designs a proactive dropping policy with its bi-direction request dropping and adaptive request priority}{\sysname designs a proactive dropping policy and adaptive request priority}, which is orthogonal to existing scheduling methods like resource scaling and dynamic batching. It could further enhance the goodput of existing systems.
    
    \item \revise{Evaluation shows that \sysname achieves $21\%$--$176\%$ higher goodput than the state-of-the-art inference systems, while reducing the drop rate and wasted computation resources by $1.6\times$--$16.7\times$ and $1.5\times$--$62\times$ respectively.}{Evaluation shows that \sysname improves goodput by $16\%$--$176\%$ over state-of-the-art systems, while reducing the drop rate and wasted computation by $1.6\times$--$17\times$ and $1.5\times$--$62\times$ respectively.} The ablation study quantifies the impact of \sysname's design choices on goodput enhancement.
\end{itemize}


%% file: content/2_background.tex
\section{Background}
\label{sec:background}

\parab{Real-time Inference Pipeline.} 
\revise{
    A typical DNN inference pipeline is designed to deliver real-time inference, providing timely response to the clients. Latency service-level objectives (SLOs) define the acceptable request latency for the client of these pipelines [cite]. Any request that fails to meet latency SLO is considered invalid. Existing inference systems employ various techniques [cite], including dynamic batching, resource scaling and GPU scheduling, to satisfy the latency objectives. However, despite implementing optimization techniques, it is impossible to guarantee that \textit{all} requests for a real-time inference pipeline can satisfy the latency objective. Unforeseen events such as workload bursts or machine failures [cite] pose challenges. For instance, during bursty workloads, computation resources cannot scale up immediately due to model cold start [cite], leading to request accumulation and eventual latency violations.
}{%
    A typical DNN inference pipeline aims to deliver real-time response to the clients, where any request missing its latency SLO is invalid. Existing systems adopt techniques such as dynamic batching, resource scaling and GPU scheduling~\cite{zhang2020enabling, zhang2023octopus, xia2023towards, fu2023autoscratch, olston2017tensorflow}, to satisfy the latency objectives. However, even with such optimizations, it is impossible to guarantee that \textit{all} requests can satisfy the latency SLO. Unforeseen events such as workload bursts or machine failures~\cite{zhang2019mark, gunasekaran2022cocktail, dean2013tail} pose challenges. For instance, during workload bursts, resources cannot scale up instantly due to model cold starts~\cite{zhang2019mark, gujarati2020serving, romero2021infaas}, causing request accumulation and latency violations.
}

\response{none}{Shorten the background to reserve space for \secref{sec:eval}-\secref{sec:discussion}.}

\begin{figure*}[t]
    \centering
    \subfloat[Goodput comparison]{
        \includegraphics[width=0.225\textwidth]{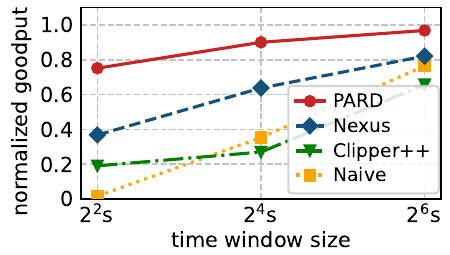}
        \label{fig:motivation_data_overall_1}
    }\hspace{0pt}
    \subfloat[Drop rate comparison]{
        \includegraphics[width=0.225\textwidth]{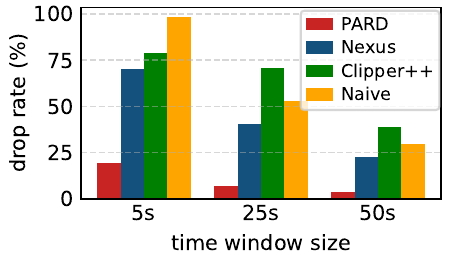}
        \label{fig:motivation_data_overall_2}
    }\hspace{0pt}
    \subfloat[Percent of dropped requests]{
        \includegraphics[width=0.225\textwidth]{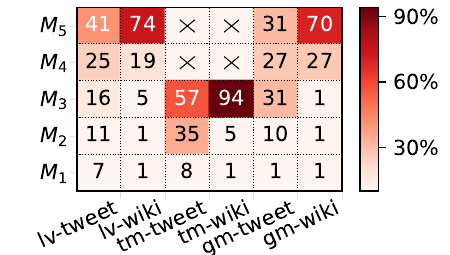}
        \label{fig:motivation_data_dropat}
    }\hspace{0pt} 
    \subfloat[Transient drop rate]{
        \includegraphics[width=0.225\textwidth]{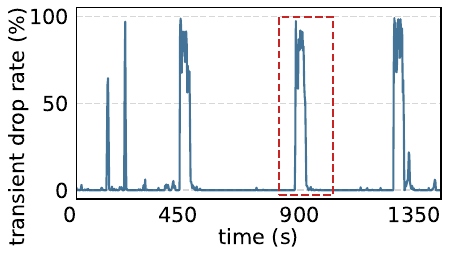}
        \label{fig:motivation_data_transient}
    }
    \caption[]{(a) and (b) The minimum goodput and corresponding drop rate across various time window sizes of existing inference systems, naive baseline, and \sysname under \texttt{lv-tweet} workload. (c) The percentage of dropped requests at each module under different workloads from \secref{sec:eval_methodology} with the reactive dropping policy. (d) Transient drop rate of the reactive dropping policy\footnotemark.}
    \label{fig:motivation_data}
    \Description{}
\end{figure*}


\begin{figure}[t]
    \centering
    \includegraphics[width=0.47\textwidth]{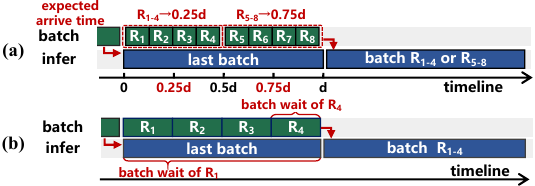}
    \caption{\revise{}{(a) The reactive dropping policy makes decisions based on request arrival order, leading to the drop-wrong-set issue. (b) Batched requests have different batch wait times $W$, ranging from $0$ to the batch execution duration $d$.}}\narrow{}
    \label{fig:motivation_example}
    \Description{}
\end{figure}

\parab{Request Dropping Policies.} To avoid the negative impact of timeout requests on goodput, recent systems adopt request dropping~\cite{shen2019nexus, gujarati2020serving, yu2022orloj, yang2022infless, ghafouri2023ipa, crankshaw2017clipper, zhang2023shepherd}. The idea is to drop requests that would miss the latency objective, especially when incoming workload exceeds system capacity. Request dropping aims to reduce the queueing delay for the remaining requests, so the remaining ones are more likely to achieve lower latency and meet the latency objective. This, in turn, enhances the overall goodput for the entire workload.


Dropping policies in existing systems fall into two types. When making dropping decisions at a given module in the pipeline, the first type drops requests that already exceed the latency objective~\cite{ghafouri2023ipa, crankshaw2017clipper}, while the second type considers both the accumulated latency in preceding modules and the inference duration of current module, and then drops requests that cannot complete within the latency objective~\cite{shen2019nexus, yu2022orloj, gujarati2020serving}. Depending on specific policy, the system drops requests that have exceeded the latency objective or have a remaining latency budget less than inference duration before inference~\cite{shen2019nexus, hu2021scrooge, ghafouri2023ipa, gujarati2020serving, crankshaw2017clipper, crankshaw2020inferline}. Since these policies only drop requests when they have to be dropped, we define them as \textit{reactive} policies. We argue that such reactive policies cannot make proper dropping decisions, leading to sub-optimal goodput, as we discuss next.

%% file: content/3_motivation.tex
\section{Motivation} \label{sec:motivation}

This section reveals the limitations of existing dropping policies (\secref{sec:motivation_limitation}) and provides in-depth analysis from two dimensions: timing of decision and criteria for selection (\secref{sec:motivation_analysis}). Finally, we identify two key questions and corresponding challenges to design an optimal dropping policy (\secref{sec:motivation_implication}).

\footnotetext{\revise{}{\figref{fig:motivation_data_dropat} and \figref{fig:motivation_data_transient} present the results of Clipper++; Nexus exhibits similar trends (\secref{sec:eval_overall}).}}

\subsection{Limitation of Reactive Dropping Policy}
\label{sec:motivation_limitation}

To highlight the limitations of current inference systems for multi-DNN inference pipelines, we evaluate Nexus~\cite{shen2019nexus} and Clipper++, a modified version of Clipper~\cite{crankshaw2017clipper} for inference pipeline workload, using the \texttt{lv-tweet} workload as described in \secref{sec:eval_methodology}. We compare them with a naive baseline that serves the inference pipeline without dropping any requests. The primary evaluation metric is \textit{goodput} and \textit{drop rate} (\ie the percentage of dropped requests to all requests).

We calculate the minimum goodput and corresponding drop rate over the entire runtime across various time window sizes in \figref{fig:motivation_data_overall_1} and \figref{fig:motivation_data_overall_2} for comparison. From the results, it is notable that the goodput of Nexus and Clipper++ may decrease to as low as $30\%$ and $21\%$ of the input request rate, with drop rates of $70\%$ and $79\%$, which performs worse than the naive baseline at several window sizes.

\subsection{Understand Why Reactive Policy Fails} 
\label{sec:motivation_analysis}

\parab{Observation \#1: Reactive policy drops requests too late in the inference pipeline.} At first glance, request dropping seems to adversely affect goodput. So existing systems drop requests only when they have already violated or are about to violate the latency objective. This reactive approach can \textit{not} make timely dropping decisions in the inference pipeline, as it only considers the accumulated latency for preceding and current modules, which drops requests too late.

\label{page:motivation_analysis}
\figref{fig:motivation_data_dropat} shows the percentage of dropped requests at each module for various workloads as described in \secref{sec:eval_methodology} under reactive dropping policy, where $57.1\%$ to $97.2\%$ of dropped requests are dropped in the latter half of the pipeline. To understand why, we take the traffic monitoring application \texttt{tm} with 3 modules cascaded (the third and fourth column of \figref{fig:motivation_data_dropat}) as an example, where $57.1\%$ and $94.4\%$ of dropped requests are dropped in the last module under two traces. 
\revise{
    As shown in \figref{fig:motivation_example_late}, if a request $R_1$ experiences long wait time [footnote: The wait time includes queuing delay and batch wait time, which will be formally introduced in \secref{sec:design_bidirectional}.] at module $M_1$ and $M_2$ due to a workload burst, it over-consumes the end-to-end latency budget. The reactive policy will \textit{not} drop $R_1$ until $M_3$, since the remaining latency budget is insufficient for $M_3$. Once dropped at $M_3$, the computation resources consumed by $R_1$ at $M_1$ and $M_2$ are wasted, leading to the \textbf{drop-too-late} issue, which increases the invalid rate [footnote: The invalid rate refers to the ratio of wasted computation resources, which will be formally defined in \secref{sec:eval_methodology}]. These wasted computations bring backpressure for preceding modules, increasing queuing delay and eventually leading to a lower goodput.
}{%
    As shown in \figref{fig:intro_dropping}a and \figref{fig:intro_dropping}b, if a request $R_1$ experiences long wait time (including queueing delay and batch wait time) in the early modules, it will over-consume the end-to-end latency budget. Existing reactive policy will \textit{not} drop $R_1$ until after the last module’s inference, when the remaining budget is already insufficient. Once dropped, the computation resources consumed by $R_1$ are wasted, causing the \textbf{drop-too-late} issue and raising the invalid rate\footnote{\revise{}{Invalid rate measures the ratio of GPU time consumed by dropped requests (\ie wasted computation resources), and it will be formally defined in \secref{sec:eval_methodology}.}}. These wasted computations bring backpressure for preceding modules, increasing queuing delay and ultimately lowering goodput.
}
\response{D}{Remove the original Figure 3a and revise the analysis with \figref{fig:intro_dropping}b.}

\parab{Observation \#2: Reactive policy drops the wrong set of requests.} Existing systems maintain a FIFO request queue for each worker, making dropping decisions based on arrival order. This arrival-order-based method can \textit{not} identify the proper set of requests to drop and over-consume latency budget at the early stages of the pipeline, leaving insufficient budget for later stages and ultimately reducing goodput.

\revise{
    \figref{fig:motivation_data_transient} shows the drop rate of reactive dropping policy under the \texttt{lv-tweet} workload with five modules cascaded, where the transient drop rate exceeds $95\%$ around $t=850s$, even though the input request rate is only doubled at the given moment (\figref{fig:eval_realtime_goodput2_norm}e). \figref{fig:motivation_example_set} illustrates the reason behind this issue, where the request input rate of $8$ req/s exceeds the system capability of $4$ req/s, and half of the requests need to be dropped to avoid accumulation. The batch size and inference duration are $4$ and $d$, respectively. 
    Based on the arrival order, the reactive policy of Nexus keeps requests $R_{1\to 4}$ for inference and drops $R_{5\to 8}$, leading to an average request wait time of around $(\frac{7.5}{8}+\frac{6.5}{8}+\frac{5.5}{8}+\frac{4.5}{8})d/4 = 0.75d$. On the contrary, keeping $R_{5\to 8}$ and dropping $R_{1\to 4}$ instead would reduce the wait time to $(\frac{3.5}{8}+\frac{2.5}{8}+\frac{1.5}{8}+\frac{0.5}{8})d/4 = 0.25d$. Thus, the reactive policy over-consumes the latency budget due to the \textbf{drop-wrong-set} issue, causing requests to violate latency objectives and increasing the drop rate.
}{%
    \figref{fig:motivation_data_transient} shows the drop rate of the reactive dropping under the \texttt{lv-tweet} workload with five cascaded modules: the transient drop rate exceeds $95\%$ around $t{=}850$s even though the input request rate only doubles at that moment (\figref{fig:eval_realtime_goodput2_norm}e). \figref{fig:motivation_example}a explains why: within a window of duration $d$ (the inference time of one batch) with batch size $4$, there are $8$ arrivals but the system can serve $4$ within the latency constraint, so exactly half must be dropped to prevent accumulation. The reactive policy serves requests in arrival order, keeping the earliest arrived $R_{1\to 4}$ and dropping $R_{5\to 8}$. In the depicted window, requests $R_{1\to 4}$ fall in $[0, 0.5d]$ with a mean arrival time of $0.25d$, so their expected batch wait time to the next batch start is $d-0.25d=0.75d$, whereas the expected batch wait time of $R_{5\to 8}$ is only $d-0.75d=0.25d$. Therefore, the reactive policy's FIFO request queue wrongly keeps $R_{1\to 4}$ with a higher batch wait time, which over-consumes the request's latency budget. This \textbf{drop-wrong-set} issue eventually causes requests to violate latency objectives and increases the drop rate.
}

\response{B/C/D}{Redesign the current \figref{fig:motivation_example}a and revise the description for clarity.}

\subsection{Implications}
\label{sec:motivation_implication}

The experimental results highlight that applying a reactive dropping policy cannot yield high goodput. To overcome the limitations of existing inference systems and ensure high goodput for inference pipelines, \sysname should adopt a \textit{proactive} dropping policy, addressing two key questions: (1) \textit{When to drop requests.} \sysname should proactively identify which requests can not be completed within the latency objective, enabling timely dropping decisions. (2) \textit{Which specific requests to drop.} \sysname should proactively select the set of requests to be dropped that enhance goodput.

However, designing such a system presents several challenges: (1) \textit{Estimating the end-to-end latency of each request:} Addressing the first question requires accurately estimating the latency of each request at each module. This is challenging due to the latency uncertainty caused by batching (\revise{\figref{fig:motivation_batching}}{\figref{fig:motivation_example}b}) and queueing. (2) \textit{Identifying main reasons for request dropping:} Addressing the second question requires the system to identify the main reasons for dropping. Given the variability and burstiness of inference workloads, these reasons may differ significantly. Next, we will introduce how \sysname addresses these two critical challenges through \revise{bi-directional}{proactive} request dropping and adaptive request priority.

%% file: content/4_design.tex
\section{System Design}
\label{sec:design}

We start with an overview of \sysname (\secref{sec:design_overview}), followed by its two key designs, proactive request dropping (\secref{sec:design_bidirectional}) and adaptive request priority (\secref{sec:design_priority}).

\subsection{System Overview}
\label{sec:design_overview} 

\begin{figure}[t]
    \centering
    \includegraphics[width=0.46\textwidth]{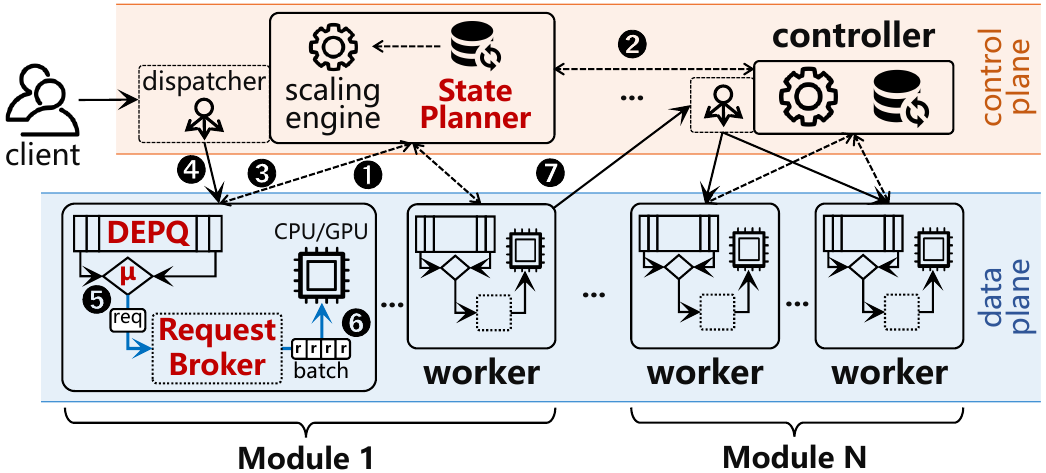}
    \caption{\sysname overview}\narrow{}\narrow{}
    \label{fig:design_overview}
    \Description{}
\end{figure}


We design \sysname, a pipeline inference system that ensures high goodput by \textit{proactively identifying requests with insufficient latency budgets and dynamically selecting proper set of requests to drop under varying workloads}. \figref{fig:design_overview} illustrates system architecture with two key components: (1) distributed controllers for request dispatching, state synchronization, and latency estimation at each module, and (2) parallel workers that execute dropping decisions for individual requests.

\sysname allocates each module in the inference pipeline one controller and multiple workers. Each controller's State Planner monitors the runtime state of each worker, including queueing delay, batch size, and throughput \ding{172}, and synchronizes these states across modules \ding{173}. Given these states, the State Planner calculates the module's required latency budget and sends it to its assigned workers \ding{174}. At runtime, requests will be dispatched among workers by dispatcher \ding{175}. Then, requests will enter a request queue called DEPQ to determine its dropping priority based on latency and workload information \ding{176}. Request Broker in each worker fetches requests from DEPQ and decides whether to drop or infer each request \ding{177}. Finally, for those inferred requests, workers redirect them to \revise{the next module}{subsequent modules} in the inference pipeline \ding{178}.

\subsection{\revise{Bi-Directional}{Proactive} Request Dropping} \label{sec:design_bidirectional}

\begin{figure*}[t]
    \centering
    \includegraphics[width=0.9\textwidth]{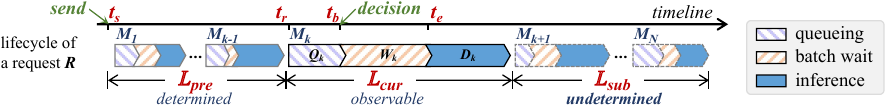}
    \caption{Lifecycle of a request $R$ sent at $t_s$ by the client in an $N$-module pipeline, where $t_r$, $t_b$, $t_e$ represent the moments when the request is received by module $M_k$, put into a batch, and when the batch execution starts, respectively. At $t_b$, \sysname could get all bi-directional runtime information for each request and make a timely dropping decision before it enters a batch.}\narrow{}
    \label{fig:design_timeline}
    \Description{}
\end{figure*}


\parab{Insight and Approach.} \textit{To timely drop requests for goodput enhancement, \sysname proactively drops requests by estimating their latency with bi-directional runtime information.} Specifically, when making dropping decisions at workers of each module, \sysname estimates the end-to-end request latency based on the request cumulative latency information from the preceding modules, as well as the batch wait time distribution and queueing latency from the subsequent modules.

\label{page:design_bidirectional}
\response{D}{Tighten the presentation of \secref{sec:design_bidirectional} for conciseness and reserving space for \secref{sec:eval}-\secref{sec:discussion}.}

\parab{Latency Decoupling.}
\revise{
    As shown in \figref{fig:design_timeline}, for a given request at module $M_k,~k \in [1, N]$ of an $N$-module inference pipeline, its end-to-end latency $\mathbb{L}$ can be decoupled into three parts: cumulative latency in preceding modules $\mathbb{L}_{pre}$, latency budget for current module $\mathbb{L}_{cur}$ and latency budget for subsequent modules $\mathbb{L}_{sub}$. Among these, only $\mathbb{L}_{pre}$ is determined at this stage. Representing the request latency in each module as $Lat_{(k)}$, the end-to-end latency $\mathbb{L}$ can be expressed as follows.
}{%
    As shown in \figref{fig:design_timeline}, for a request at module $M_k,~k \in [1, N]$ of an $N$-module inference pipeline, its end-to-end latency $\mathbb{L}$ can be decoupled into three parts: cumulative latency from preceding modules $\mathbb{L}_{pre}$, current module’s latency budget $\mathbb{L}_{cur}$, and latency budget for subsequent modules $\mathbb{L}_{sub}$. At this stage, only $\mathbb{L}_{pre}$ is determined. If latency in module $k$ is $Lat_{(k)}$, then $\mathbb{L}$ can be expressed as:
}
{\abovedisplayskip=3pt
\belowdisplayskip=3pt
\begin{subequations}\label{eqn:design_latency_e2e}
\begin{align}
    \label{eqn:design_latency_e2e_a}
    \mathbb{L}~ &= ~~~{\mathbb{L}}_{pre} ~~~+~~ {\mathbb{L}}_{cur} ~+~~~~ {\mathbb{L}}_{sub} \\
               &= 
    \textstyle\sum_{i=1}^{k-1}Lat_{(i)} + Lat_{(k)} + \sum_{i=k+1}^{N}Lat_{(i)}
\end{align}
\end{subequations}}

\revise{
    By decoupling the end-to-end latency $\mathbb{L}$, we can re-examine why reactive policy fails and how \sysname should make dropping decisions. The reactive policy considers only $\mathbb{L}_{pre}$ and $\mathbb{L}_{cur}$ in \eqnref{eqn:design_latency_e2e_a} for dropping decisions, ignoring the latency budget for subsequent modules ($\mathbb{L}_{sub}$). This omission prevents timely detection of latency violations, causing incorrect decisions. So a proper policy should proactively drop requests based on end-to-end latency for timely decisions.
}{%
    This decoupling clarifies why reactive policy fails and how \sysname should make dropping decisions. Reactive approaches considers only $\mathbb{L}_{pre}$ and $\mathbb{L}_{cur}$ in \eqnref{eqn:design_latency_e2e_a}, ignoring the latency budget for subsequent modules ($\mathbb{L}_{sub}$). This omission prevents timely detection of latency violations, causing incorrect decisions. A proper policy should proactively drop requests based on end-to-end latency for timely decisions.
}

\revise{
    However, making the proper dropping decisions is challenging due to the uncertainty of request latency $Lat_{(k)}$ in each subsequent module. As the batching process shown in \figref{fig:motivation_batching}, when a request arrives at module $M_k$, it enters the request queue and begins waiting. The scheduler collects the next batch when the previous batch has just started inference to prevent GPU idling. Consequently, $Lat_{(k)}$ can be decoupled into three components: (1) queueing delay $Q_k$, the time before a request is collected to a batch; (2) batch wait time $W_k$, the time between a request is collected to a batch and the start of inference; and (3) model execution duration $D_k$. Then, $Lat_{(k)}$ can be expressed as \eqnref{eqn:design_latency_module}.
}{%
    However, making correct dropping decisions is challenging due to the uncertainty of request latency $Lat_{(k)}$ in subsequent modules. As shown in \figref{fig:motivation_example}b, when a request arrives at module $M_k$, it enters the request queue and waits. The scheduler collects the next batch right after the previous one begins execution to avoid GPU idling. Thus, $Lat_{(k)}$ consists of three components: (1) queueing delay $Q_k$, the time before a request is added to a batch; (2) batch wait time $W_k$, the time between being added to a batch and the start of inference; and (3) execution duration $D_k$. Formally:
}
{\abovedisplayskip=3pt
\belowdisplayskip=3pt
\begin{equation}\label{eqn:design_latency_module}
    Lat_{(k)} = Q_k + W_k+ D_k, ~k \in [1, N]
\end{equation}}

\revise{
    As shown in \figref{fig:motivation_batching}, the batch wait time of each request is uncertain and distributed between $0$ and $M_k$'s duration $d_k$, depending on the time the request is collected into the batch. This uncertainty makes the aggregated batch wait time $\sum_{i=k+1}^{N}W_i$, which is part of $\mathbb{L}_{sub}$ in \eqnref{eqn:design_latency_e2e_a}, unpredictable and exhibits a huge variance between $0$ and $\sum_{i=k+1}^{N}d_i$. Similarly, each request's aggregated queueing delay $\sum_{i=k+1}^{N}Q_i$ varies dynamically with the workload. Over-estimating them leads to excessive request dropping, while under-estimating them will also result in a higher invalid rate, both of which eventually lead to a lower goodput (\secref{sec:eval_ablation}). 
}{%
    As shown in \figref{fig:motivation_example}b, batch wait time is uncertain and varies between $0$ and the execution duration $d_k$, depending on when the request is added to the batch. This makes the aggregated batch wait time $\sum_{i=k+1}^{N} W_i$, a component of $\mathbb{L}_{sub}$ in \eqnref{eqn:design_latency_e2e_a}, highly unpredictable and ranges from $0$ to $\sum_{i=k+1}^{N} d_i$. Similarly, aggregated queueing delay $\sum_{i=k+1}^{N} Q_i$ fluctuates with workload dynamics. Overestimating these values leads to excessive dropping, while underestimating them increases invalid requests, both ultimately reducing goodput (\secref{sec:eval_ablation}).
}

In summary, to make timely dropping decisions, it's critical to properly estimate the latency for subsequent modules and the one for the preceding and current modules. Next, we introduce how \sysname estimates the end-to-end latency $\mathbb{L}$ via its novel State Planner and Request Broker to decide when to drop with bi-directional runtime information.

\parab{State Planner.}
\revise{
    Each module's State Planner in \sysname makes a proper estimation of unpredictable request latency for subsequent modules $\mathbb{L}_{sub}$. As mentioned above, $\sum_{i=k+1}^{N}Q_i$, $\sum_{i=k+1}^{N}W_i$, and $\sum_{i=k+1}^{N}D_i$ have different physical implications and properties, and any one of them may become a major component of $\mathbb{L}$ in certain cases (\eg a sharp increase in $\sum_{i=k+1}^{N}Q_i$ due to request burst). Therefore, \sysname estimates the three components independently as follows:
}{%
    In \sysname, each module’s State Planner estimates the unpredictable request latency of subsequent modules $\mathbb{L}_{sub}$. As discussed, $\sum_{i=k+1}^{N} Q_i$, $\sum_{i=k+1}^{N} W_i$, and $\sum_{i=k+1}^{N} D_i$ have distinct physical implications and characteristics, and any of them can dominate $\mathbb{L}$ under certain conditions (\eg a surge in $\sum_{i=k+1}^{N} Q_i$ due to bursts). Thus, \sysname estimates these three components independently:
}

\begin{itemize}[leftmargin=12pt,labelsep=6pt,topsep=0pt,itemsep=0pt]
    \revise{
        \item  $\sum_{i=k+1}^{N}Q_i$: State Planner of each module monitors recent average queueing delay $q_i$ via sliding window, synchronizing with other modules. Then it calculates the cumulative queueing delay with synchronized information from subsequent modules as $\sum_{i=k+1}^{N}Q_i = \sum_{i=k+1}^{N}q_i$.
        
        \item $\sum_{i=k+1}^{N}D_i$: State Planner periodically synchronize the batch size of each module in the inference pipeline, and then calculates the cumulative execution duration of subsequent modules, according to offline model profiling~\cite{shen2019nexus, choi2021lazy}. Namely, we have $\sum_{i=k+1}^{N}D_i = \sum_{i=k+1}^{N}d_i$, where $d_i$ is the profiled execution duration.
        
        \item $\sum_{i=k+1}^{N}W_i$: As shown in \figref{fig:motivation_batching}, for each request, the batch wait time $W_i$ of module $M_i$ exhibits uncertainty and is distributed between $[0, d_i]$. Further, this uncertainty will be amplified for the entire inference pipeline as modules are cascaded, leading to an unpredictable aggregated batch wait time $\sum_{i=k+1}^{N}W_i$ ranging from $0$ to $\sum_{i=k+1}^{N}d_i$ at module $M_k$. However, to make timely and correct dropping decision for each specific request, its aggregated batch wait time must be properly estimated. Next, we introduce how \sysname derives it.
    }{%
        \item  $\sum_{i=k+1}^{N} Q_i$: Each State Planner monitors the recent average queueing delay $q_i$ using a sliding window\footnote{\revise{}{Default to a $5$s linear weighted window, with sensitivity analysis in \secref{sec:eval_sensitivity}.}} and synchronizes with other modules. The cumulative queueing delay is then calculated as $\sum_{i=k+1}^{N} Q_i = \sum_{i=k+1}^{N} q_i$.

        \item $\sum_{i=k+1}^{N} D_i$: The State Planner periodically synchronizes batch sizes across modules and computes cumulative execution duration using offline model profiling~\cite{shen2019nexus, choi2021lazy}, \ie $\sum_{i=k+1}^{N} D_i = \sum_{i=k+1}^{N} d_i$, where $d_i$ is the profiled duration.
        
        \item $\sum_{i=k+1}^{N} W_i$: As shown in \figref{fig:motivation_example}b, the batch wait time $W_i$ at module $M_i$ varies between $0$ and $d_i$, depending on when a request enters the batch. This uncertainty accumulates across cascaded modules, making $\sum_{i=k+1}^{N} W_i$ unpredictable, ranging from $0$ to $\sum_{i=k+1}^{N} d_i$ at module $M_k$. For timely and accurate dropping decisions, this aggregated wait time must be carefully estimated. We next explain how \sysname derives it.
    }
\end{itemize}

\parae{Batch wait estimation.}
\revise{
    The exact value of $\sum_{i=k+1}^{N}W_i$ for each request is unpredictable and varies widely, leading to under- or over-estimation issues. However, the distribution of $\sum_{i=k+1}^{N}W_i$ can be derived during runtime, allowing us to derive a balanced estimation of $\sum_{i=k+1}^{N}W_i$, denoted as $w_k$, when making dropping decisions at module $M_k$.
}{%
    The exact value of $\sum_{i=k+1}^{N} W_i$ for each request is highly variable and unpredictable, leading to under- or over-estimation issues. Nevertheless, its runtime distribution can be observed, allowing \sysname to derive a balanced estimate, denoted as $w_k$, when making dropping decisions at module $M_k$.
}

\revise{
    First, under- and over-estimation of batch wait time reduce goodput for two opposite reasons. Under-estimating $\sum_{i=k+1}^{N}W_i$ (\eg $w_k=0$) results in fewer requests being dropped in the current module due to aggregated batch wait time estimation\footnote{Requests can still be dropped due to other factors such as long queueing delay in preceding modules.}. However, these requests may later be dropped in subsequent modules due to higher actual batch wait times than the estimation. This increases the invalid rate and backpressure on preceding modules, leading to a higher overall drop rate. Conversely, over-estimating $\sum_{i=k+1}^{N}W_i$ (\eg $w_k = \sum_{i=k+1}^{N}d_i$) results in many requests being mistakenly dropped at the current module due to strict batch wait time constraint for subsequent modules, also leading to a higher drop rate. Thus, \sysname aims to identify a \textit{sweet spot} $w_k$ for each module $M_k$ within the range $[0, \sum_{i=k+1}^{N}d_i]$ to balance invalid and drop rates, thereby enhancing overall goodput.
}{%
    First, both under- and over-estimation reduce goodput, but for opposite reasons. Under-estimating $\sum_{i=k+1}^{N} W_i$ (\eg $w_k=0$) causes fewer requests to be dropped at the current module\footnote{Requests may still be dropped due to other factors, such as long queueing delay in preceding modules.}. These \textit{mis-kept} requests are likely to be dropped later in subsequent modules when the actual batch wait time exceeds the estimate, increasing the invalid rate and backpressure, which ultimately raises the overall drop rate. Conversely, over-estimating $\sum_{i=k+1}^{N} W_i$ (\eg $w_k=\sum_{i=k+1}^{N} d_i$) leads to premature drops at the current module, since the assumed wait time constraint for downstream modules is overly strict. These \textit{mis-dropped} requests directly increase the drop rate. To avoid both extremes, \sysname seeks a \textit{sweet spot} $w_k$ within $[0, \sum_{i=k+1}^{N} d_i]$ that balances invalid and drop rates, thereby improving goodput.
}

\begin{figure}[!t]
    \centering
    \includegraphics[width=0.485\textwidth]{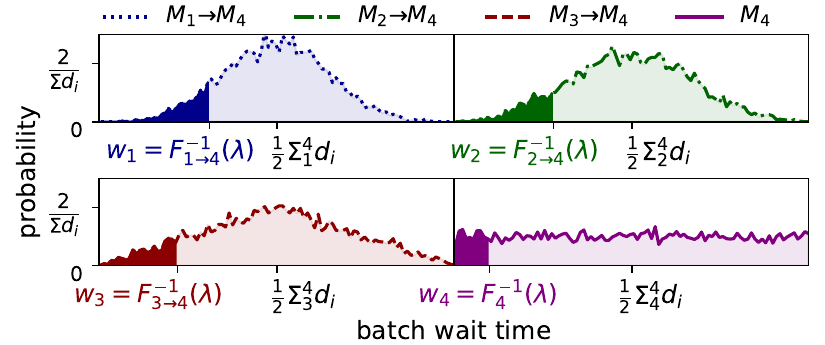}
    \caption{Probability density of requests' total batch wait time at each module in a 4-modules inference pipeline.}\narrow{}
    \label{fig:design_distribution}
    \Description{}
\end{figure}

\revise{
    Second, the optimal $w_k$ depends on the module's position in pipeline. \figref{fig:design_distribution} shows the probability density of the aggregated batch wait time for a pipeline with four cascaded modules and fixed batch sizes, based on sampling $10k$ requests from the \texttt{wiki} trace described in \secref{sec:eval_methodology}. As the number of cascaded modules increases, the aggregated batch wait time becomes more concentrated around $\frac{1}{2}\sum_{i=k+1}^{N}d_i$. For instance, the aggregated batch wait time from $M_1$ to $M_4$ is mainly centred around $\frac{1}{2}\sum_{i=1}^{4}d_i$, while $M_4$'s aggregated batch wait time for itself is evenly distributed from $0$ to $d_4$. Thus, to ensure effective constraints for all modules, $w_k$ should be closer to $\frac{1}{2}\sum_{i=k+1}^{N}d_i$ if there are more modules after $M_k$.
}{%
    Second, the optimal $w_k$ depends on the module’s position in the pipeline. \figref{fig:design_distribution} shows the probability density of aggregated batch wait time in a four-module pipeline with fixed batch sizes, based on $10k$ sampled requests from the \texttt{wiki} trace (\secref{sec:eval_methodology}). As more modules are cascaded, the aggregated batch wait time of weakly correlated $W_i$ becomes concentrated around $\frac{1}{2}\sum_{i=k+1}^{N} d_i$, following the central limit theorem. For example, the total wait time from $M_1$ to $M_4$ is centered near $\frac{1}{2}\sum_{i=1}^{4} d_i$, whereas $M_4$’s own wait time is uniformly distributed over $[0, d_4]$. Hence, for modules earlier in the pipeline, $w_k$ should be set closer to $\frac{1}{2}\sum_{i=k+1}^{N} d_i$ to provide effective constraints across the pipeline.
}

\response{D}{Clarify that $\sum_{i=k+1}^{N} W_i$ approaches $\frac{1}{2}\sum_{i=k+1}^{N} d_i$ due to the central limit theorem.}

\revise{
    Based on the observations above, State Planner employs a three-round heuristic to determine an appropriate $w_k$ for each module. First, it performs random sampling\footnote{The runtime complexity is $O(M(N-k+1))$, where $M$ is the length of the collected arrival process with a default value of $10,000$.} on the collected arrival process from recent workload to derive the probability density function (PDF) $F_{k+1\rightarrow N}$ for each module's aggregated batch wait time from $M_{k+1}$ to $M_N$. Second, it defines $\lambda$, where $\lambda \in [0,1]$, as the quantile used to determine $w_k$ such that $F_{k+1\rightarrow N} = \lambda$. This means $\lambda$ proportion of requests are expected to have an aggregated batch wait time no larger than $w_k$. Third, it estimates the aggregated batch wait time as $w_k = F_{k+1\rightarrow N}^{-1}(\lambda)$ for each module in the inference pipeline. Thus, each module's Request Broker can make a dropping decision using $\sum_{i=k+1}^{N}W_i = w_k$.
}{%
    Based on the above observations, the State Planner applies a three-round heuristic to determine $w_k$ for each module. First, it performs random sampling\footnote{The runtime complexity is $O(M(N-k+1))$, where $M$ is the length of the collected arrival process, default $M{=}10{,}000$.} on recent arrivals to derive the probability density function (PDF) $F_{k+1\rightarrow N}$ for aggregated batch wait time from $M_{k+1}$ to $M_N$. Second, it selects a quantile $\lambda \in [0,1]$ such that $F_{k+1\rightarrow N}=\lambda$, meaning that $\lambda$ proportion of requests have aggregated wait times no greater than $w_k$. Third, it estimates $w_k = F_{k+1\rightarrow N}^{-1}(\lambda)$, which is then used by each module’s Request Broker for dropping decisions via $\sum_{i=k+1}^{N} W_i = w_k$.
}

\revise{
    Intuitively, $\lambda$ serves as a control knob on how aggressively the \sysname should estimate $\sum_{i=k+1}^{N}W_i$ for subsequent modules. When $\lambda=0$, $w_k=F^{-1}(0)=0$, indicating the lower bound estimation. Conversely, when $\lambda=1$, $w_k=F^{-1}(1)=\sum_{i=k+1}^{N}d_i$, indicating the upper bound. Empirically, $\lambda$ is set to $0.1$ by default, and we conduct a sensitivity analysis in \secref{sec:eval_ablation}. With the quantile $\lambda$, \sysname could estimate $w_k$ for each module that satisfies two properties: First, $w_k$ is closer to $\frac{1}{2}\sum_{i=k+1}^{N}d_i$ if there are more modules after $M_k$. Second, $w_k$ varies in real-time as the batch size of each module changes. For example, in a four-module inference pipeline (\figref{fig:design_distribution}), suppose every module's execution duration is $d$ at the given moment, the quantile $\lambda = 0.1$ allows State Planner calculate $w_k$ as follows: 
}{%
    Intuitively, $\lambda$ serves as a control knob on how aggressively \sysname estimates $\sum_{i=k+1}^{N} W_i$ for subsequent modules. When $\lambda=0$, $w_k=F^{-1}(0)=0$ (lower bound); when $\lambda=1$, $w_k=F^{-1}(1)=\sum_{i=k+1}^{N} d_i$ (upper bound). Empirically, $\lambda$ is set to $0.1$ by default, and sensitivity analysis is presented in \secref{sec:eval_ablation}. With this quantile, \sysname estimates $w_k$ with two desirable properties: (1) $w_k$ approaches $\frac{1}{2}\sum_{i=k+1}^{N} d_i$ as the number of subsequent modules increases; and (2) $w_k$ varies in real time as batch sizes changes. For example, in a four-module pipeline (\figref{fig:design_distribution}) with equal duration $d$, $\lambda=0.1$ yields:
}
{\abovedisplayskip=3pt
\belowdisplayskip=3pt
\begin{equation}
\begin{aligned}
    w_1=\textbf{0.31}{\textstyle\sum\nolimits_1^4} d_i = 1.24d \quad & 
    w_2=\textbf{0.28}{\textstyle\sum\nolimits_2^4} d_i = 0.84d \\
    w_3=\textbf{0.22}{\textstyle\sum\nolimits_3^4} d_i = 0.44d \quad &
    w_4=\textbf{0.10}{\textstyle\sum\nolimits_4^4} d_i = 0.10d
    \nonumber
\end{aligned}
\end{equation}}

\revise{
    With such estimation, \sysname may still mis-keep or mis-drop requests. However, it optimizes $w_k$ to balance the drop and invalid rates for goodput enhancement (\secref{sec:eval_ablation}).
}{%
    Although some requests may still be mis-kept or mis-dropped, this heuristic balances drop and invalid rates to improve goodput (\secref{sec:eval_ablation}).
}

\begin{figure}[t]
    \centering
    \includegraphics[width=0.42\textwidth]{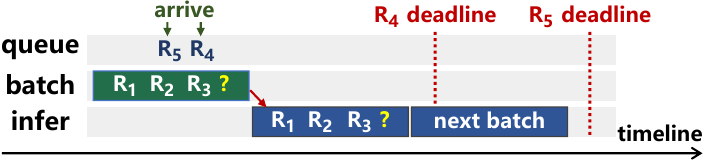}
    \caption{Example of dropping under steady workload.}\narrow{}
    \label{fig:design_priority}
    \Description{}
\end{figure}

\parab{Request Broker.} 
\revise{
    \sysname leverages Request Broker to estimate the end-to-end latency $\mathbb{L}$ according to \eqnref{eqn:design_latency_e2e} and \eqnref{eqn:design_latency_module} as follows.
}{%
    \sysname’s Request Broker estimates the end-to-end latency $\mathbb{L}$ using \eqnref{eqn:design_latency_e2e} and \eqnref{eqn:design_latency_module}:
}

\parae{Backward.}
\revise{
    As shown in \figref{fig:design_timeline}, when a request reaches module $M_k$ at timestamp $t_r$, its cumulative latency $\mathbb{L}_{pre}$ for $M_k$'s preceding modules is determined and obtainable. Specifically, it equals the duration from when the request is sent ($t_s$) to when it arrives at $M_k$ ($t_r$). Namely, we have $\mathbb{L}_{pre} = t_r - t_s$.
}{%
    As shown in \figref{fig:design_timeline}, when a request reaches module $M_k$ at timestamp $t_r$, its cumulative latency $\mathbb{L}_{pre}$ from preceding modules is already determined. Specifically, it equals the elapsed time from when the request was sent ($t_s$) to when it arrives at $M_k$ ($t_r$): $\mathbb{L}_{pre} = t_r - t_s$.
}

\parae{Current.} Request Broker leverages \eqnref{eqn:design_latency_module} to calculate the latency $\mathbb{L}_{cur}$ of module $M_k$ in three steps.

\begin{itemize}[leftmargin=12pt,labelsep=6pt,topsep=0pt,itemsep=0pt]
    \revise{
        \item $Q_k$: As shown in \figref{fig:design_timeline}, when Request Broker receives the given request from the request queue at timestamp $t_b$, its queueing delay $Q_k$ is determined as $Q_k = t_b - t_r$.
        \item $W_k$: Each worker in \sysname adopts a batch processing pipeline to avoid idle cycles on GPUs, where Request Broker begins to collect the next batch only when the previous one starts execution. Therefore, the Request Broker at each worker can accurately derive the expected starting time $t_e$ for a given batch, which equals the expected ending time for its previous batch. Then the batch wait time $W_k$ is calculated as $W_k = t_e - t_b$.
        \item $D_k$: The execution duration $D_k$ is determined by the DNN model and its batch size, which is obtained via offline profiling. Namely, we have $D_k = d_k$, where $d_k$ is the profiled execution duration at current batch size.
    }{%
        \item $Q_k$: When the request is dequeued at timestamp $t_b$, queueing delay is $Q_k = t_b - t_r$.
        \item $W_k$: Workers collect the next batch only after the previous batch begins execution to avoid GPU idling. Thus, Request Broker can derive the expected start time $t_e$ of the current batch, equal to the expected end of the previous batch. The batch wait time is then $W_k = t_e - t_b$.
        \item $D_k$: Execution duration depends on the DNN model and batch size, obtained from offline profiling, \ie $D_k = d_k$, where $d_k$ is the profiled duration at the current batch size.
    }
\end{itemize}

\parae{Forward.} As discussed, Request Broker collects the latency information for subsequent modules $\mathbb{L}_{sub} = \sum_{i=k+1}^{N}Q_i + \sum_{i=k+1}^{N}D_i + \sum_{i=k+1}^{N}W_i$ from State Planner.


In summary, as shown in \figref{fig:design_overview}, before a request enters a batch via Request Broker (\ie at $t_b$ in \figref{fig:design_timeline}), Request Broker and State Planner could obtain all bi-directional runtime information in \eqnref{eqn:design_latency_e2e} and \eqnref{eqn:design_latency_module}. Request Broker then calculates end-to-end latency $\mathbb{L}$ as in \eqnref{eqn:design_latency_done}.
\revise{}{%
    For directed acyclic graph (DAG) pipelines, where each vertex is a module, \sysname estimates the latency $\mathbb{L}'$ of a request along each subsequent DAG path and takes the maximum as the end-to-end latency estimation.
}
\response{A/B/D/E}{Clarify how to estimate request latency in DAG-style pipelines.}
\label{page:design_RAG}
{
\abovedisplayskip=0pt
\belowdisplayskip=-6pt
\begin{subequations}\label{eqn:design_latency_done}
\begin{align}
    \mathbb{L} &= {\mathbb{L}}_{pre} + {\mathbb{L}}_{cur} + {\mathbb{L}}_{sub} \\
               &= \underbrace{t_e - t_s + d_k}_{\text{Request Broker}} + 
                  \underbrace{\textstyle \sum_{i=k+1}^{N}q_i + \sum_{i=k+1}^{N}d_i + F_{k+1\rightarrow N}^{-1}(\lambda)}_{\text{State Planner}}
\end{align}
\end{subequations}
}

\subsection{Adaptive Request Priority}
\label{sec:design_priority}

\parab{Insight and Approach.} \textit{To select the set of requests to drop that enhances goodput, \sysname dynamically reorders requests based on their remaining latency budget and workload intensity.} \sysname employs a double-ended priority queue (DEPQ) to alter the request decision order and designs adaptive request prioritization to handle bursty workloads as follows.

\revise{
    First, \sysname should reorder requests based on their remaining latency budgets, as the arrival order does not determine the remaining latency. Batching introduces significant uncertainty in latency, meaning consecutively arriving requests can experience different batch wait times in preceding modules, making the arrival order ineffective for determining the remaining latency budget. Therefore, \revise{as the example in \figref{fig:motivation_example_set} shows}{as the example in \figref{fig:motivation_example}a shows}, \sysname should use an efficient reordering strategy based on remaining latency budgets to select the proper set of requests to drop.
}{%
    First, requests should be reordered by remaining latency budgets, since arrival order does not reflect them. Batching introduces uncertainty: consecutively arriving requests may experience different wait times in preceding modules, making arrival order ineffective for estimating remaining budgets. Therefore, \sysname should use an efficient reordering strategy based on remaining latency, as illustrated in \figref{fig:motivation_example}a, to select the proper set of requests to drop.
}

\revise{
    Second, \sysname should dynamically reorder requests based on the workload, as the primary reasons for request dropping vary with workloads. As aforementioned, during bursts, the primary reason for dropping is the queueing effect caused by request accumulation. In this case, \sysname's reordering should reduce the queueing delay to preserve a larger latency budget for subsequent modules. 

    Conversely, when the workload is steady (\ie the input workload does not exceed the worker throughput), the main reason for dropping is the uncertainty of request latency. For example, as shown in \figref{fig:design_priority}, consider two requests $R_4$ and $R_5$ in the queue when the worker needs one more request to form a batch. $R_4$ has experienced longer latency in preceding modules and arrives later than $R_5$. If the worker processes $R_5$ first, $R_4$ will miss the deadline due to the long batch wait time and be dropped. However, both $R_4$ and $R_5$ could meet the latency objective if $R_4$ is processed first. In this case, \sysname's reordering should reduce the queueing delay and batch wait time of requests with lower latency budgets, enabling more requests to be completed within latency objective.
}{%
    Second, \sysname should reorder requests dynamically based on workload, since the causes of request dropping vary with workloads. During bursts, drops mainly arise from queueing delays caused by request accumulation. In this case, \sysname's reordering should reduce the queueing delay, thereby preserving larger latency budget for subsequent modules. Conversely, under steady workloads (\ie input workload does not exceed worker throughput), the main reason for dropping is latency uncertainty. For example, in \figref{fig:design_priority}, suppose a worker needs one more request to form a batch, with $R_4$ and $R_5$ waiting. $R_4$ has experienced longer latency in preceding modules and arrives later than $R_5$. If $R_5$ is chosen, $R_4$ will suffer a long batch wait and miss the deadline; if $R_4$ is chosen, both meet the latency objective. In such cases, reordering should prioritize requests with lower remaining budgets, reducing queueing and wait time to complete more requests within the latency objective.

}

\revise{
    In summary, \sysname should select which requests to drop to enhance goodput across various scenarios, while adapting to highly bursty and variable inference workloads. To achieve this, \sysname designs two request prioritization mechanisms and a delayed adaptive priority transition policy, reordering requests based on real-time remaining latency budget and workload intensity to enhance goodput.
}{%
    In summary, \sysname should adapt to bursty and variable inference workloads by selectively dropping requests to improve goodput. To achieve this, it employs two request prioritization mechanisms and a delayed adaptive priority transition policy, reordering requests in real-time based on remaining latency budget and workload intensity.
}

\parab{Request Priority with DEPQ.}
\revise{
    \sysname's request prioritization mechanism decides which requests to drop based on the remaining latency budget and workload intensity. This enables \sysname to drop the correct set of requests, enhancing system goodput. \revise{We define the module load factor, $\mu$, as the ratio of input workload ($\boldsymbol{T_{in}}$) to module throughput ($\boldsymbol{T_{m}}$), indicating whether the inference system is under-provisioned.}{We define the module load factor $\mu = \boldsymbol{T_{in}}/\boldsymbol{T_{m}}$, with $\boldsymbol{T_{in}}$ representing the input workload and $\boldsymbol{T_{m}}$ the module throughput derived by batch size and execution duration, which together indicate whether the system is under-provisioned.} Based on $\mu$, we provide two types of prioritization:
}{%
    \sysname prioritizes requests using a double-ended priority queue (DEPQ), deciding which to drop based on remaining latency budgets and workload intensity. This ensures that the correct set of requests is dropped. We define the module load factor $\mu = \boldsymbol{T_{in}} / \boldsymbol{T_{m}}$, where $\boldsymbol{T_{in}}$ is the input workload and $\boldsymbol{T_{m}}$ is the module throughput determined by batch size and execution duration. $\mu$ thus indicates whether the system is under-provisioned. Depending on $\mu$, \sysname applies two policies:
}

\response{B}{Replace paraphrase math with equations.}

\begin{itemize}[leftmargin=12pt,labelsep=6pt,topsep=0pt,itemsep=0pt]
    \revise{
        \item When $\mu > 1$, the system is under-provisioned. To avoid over-consuming the latency budget due to excessive queueing, \sysname uses the High Budget First (HBF) mechanism, which prioritizes requests with the largest remaining latency budgets. This preserves more budget for subsequent modules, thereby reducing the probability of requests being dropped in later stages and leading to a lower drop rate and invalid rate (\secref{sec:eval_ablation}).
        \item When $\mu \le 1$, the workload is within module's processing capability. To avoid unnecessary drops due to latency uncertainty, \sysname uses the Low Budget First (LBF) mechanism, which prioritizes requests with the smallest remaining latency budgets. This reduces the probability of dropping requests with smaller latency budgets, leading to a lower drop rate and higher goodput (\secref{sec:eval_ablation}).
    }{%
        \item $\mu > 1$: The system is under-provisioned. To avoid excessive queueing that consumes latency budgets, \sysname applies the High Budget First (HBF) policy, prioritizing requests with the largest remaining budgets. This preserves more budget for subsequent modules, thereby reducing the probability of requests being dropped, lowering drop rate and invalid rate (\secref{sec:eval_ablation}).
        \item $\mu \le 1$: The workload is within module's processing capacity. To reduce unnecessary drops caused by latency uncertainty, \sysname applies the Low Budget First (LBF) policy, prioritizing requests with the smallest remaining budgets. This reduces the probability of dropping requests with smaller latency budgets, leading to a lower drop rate and higher goodput (\secref{sec:eval_ablation}).
    }
\end{itemize}

\sysname implements both prioritization strategies using a DEPQ in each worker, with the remaining latency budget as the priority. The DEPQ can simultaneously pop requests with the largest and smallest remaining latency budgets using a min-max heap. \sysname uses DEPQ to enable more effective dropping decisions based on remaining latency budget and workload burstiness rather than arrival order.



\parab{Adaptive Priority with Delayed Transition.} 
\revise{
    To provide a smooth transition between HBF and LBF, \sysname adaptively adjusts the prioritization mechanism using \algoref{algo:priority}, which decides DEPQ's priority with two thresholds $Th_{HBF}$ and $Th_{LBF}$. Specifically, when $\mu > Th_{HBF} = 1.0 + \epsilon$, \sysname switches the priority to HBF, and when $\mu < Th_{LBF} = 1.0 - \epsilon$, \sysname switches it to LBF. If $\mu \in [1.0 - \epsilon, 1.0 + \epsilon]$, \sysname does \textit{not} switch the priority, preventing frequent priority changes due to workload fluctuations and ensuring higher goodput (\secref{sec:eval_ablation}). The boundary $\epsilon$ is dynamically calculated based on workload fluctuations, where $\epsilon$ is defined as $\epsilon = \frac{\sum\nolimits \lvert \boldsymbol{T_{in}} - \boldsymbol{T_s} \rvert}{\sum \boldsymbol{T_{in}}}$, with $\boldsymbol{T_s}$ representing the workload sequence filtered by a sliding window average, ensuring reliable priority transitions and high goodput under various workloads (\secref{sec:eval_overall}).
}{
    To ensure smooth switching between HBF and LBF, \sysname adapts prioritization based on two thresholds, $Th_{HBF}=1.0+\epsilon$ and $Th_{LBF}=1.0-\epsilon$. When $\mu > Th_{HBF}$, it switches to HBF; when $\mu < Th_{LBF}$, it switches to LBF. For $\mu$ within $[1.0-\epsilon, 1.0+\epsilon]$, the priority remains unchanged, avoiding frequent changes from workload fluctuations and sustaining higher goodput (\secref{sec:eval_ablation}). The boundary $\epsilon$ is computed dynamically as $\epsilon = \frac{\sum \lvert \boldsymbol{T_{in}} - \boldsymbol{T_s} \rvert}{\sum \boldsymbol{T_{in}}}$, where $\boldsymbol{T_s}$ is the workload smoothed by a sliding-window average. This allows $\epsilon$ to expand under bursty workloads, suppressing frequent switches and ensuring stable priority transitions with consistently high goodput (\secref{sec:eval_overall}).

}

\response{D}{Clarify the effect of the boundary $\epsilon$.}
\response{none}{Remove previous Algorithm 1 temporarily due to page limit.}

%% file: content/5_evaluation.tex
\section{Evaluation}
\label{sec:eval}

We compare \sysname against existing inference systems in terms of its goodput and request drop rate. Then, we perform an ablation study to quantify the importance of \sysname's design decisions on goodput enhancement.

\subsection{Methodology} \label{sec:eval_methodology}

\label{page:implementation}
\parab{Implementation.} 
\revise{
    We implement all features of \sysname described in \secref{sec:design} in roughly 12k lines of Python code, utilizing PyTorch~\cite{paszke2019pytorch} as the default inference backend. Each component of \sysname in \figref{fig:design_overview} is containerized to ensure resource isolation and enable elastic scalability. Communication between containerized components is handled via gRPC to manage runtime state and inference requests efficiently.
}{%
    We implement \sysname in roughly $15$k lines of Python: $6.5$k for system and tests, $5$k for the application library, and the rest for benchmarks. PyTorch~\cite{paszke2019pytorch} serves as the default inference backend. Before startup, \sysname performs an offline profiling to obtain per-model execution duration and throughput under various batch sizes for online latency estimation. \sysname also adopts dynamic batching and resource scaling similar to~\cite{shen2019nexus, crankshaw2020inferline}: yields feasible batch sizes and per-worker throughput based on offline profiling, and adjusts the number of workers per module based on request rate and per-worker throughput at runtime. Controllers and workers run in separate containers, communicating via gRPC for low-overhead request routing and state exchange.
}

\response{B}{Update Line-of-Code and provide a breakdown.}

\revise{}{
    To support widely deployed DAG-style pipeline applications~\cite{zeng2025online, hu2021scrooge}, \sysname defines an inference pipeline via a JSON file composed of multiple module configurations $(\texttt{name}, \texttt{id}, \texttt{pres}, \texttt{subs})$, where \texttt{name} is the module registered in the application library, \texttt{pres} and \texttt{subs} specify the preceding and subsequent module IDs. \sysname automatically splits requests when \texttt{subs} contains multiple modules and merges sub-requests when \texttt{pres} has multiple inputs. In such cases, the State Planner estimates a request’s end-to-end latency as the maximum across all DAG paths, as described in \secref{sec:design_bidirectional}.
}
\response{C}{Clarify the integration of dynamic batching and resource scaling.}
\response{A/B/D/E}{Refactor \sysname to adapt DAG workloads (evaluated in \secref{sec:eval_overall}).}

\label{page:testbed}
\parab{Testbed.} 
\revise{
    All experiments are conducted on a cluster of 16 machines, each equipped with 4 NVIDIA 2080Ti GPUs and 48 CPU cores. The cluster is virtualized into 64 worker containers, each assigned a GPU. Containers in the cluster use NTP~\cite{mills2010network} to provide high-precision clock synchronization.
}{%
    Experiments run on a $16$-machine cluster, each with $4$ NVIDIA 2080Ti GPUs and $48$ CPU cores. We virtualize the cluster into $64$ worker containers (one GPU per container). NTP~\cite{mills2010network} keeps clocks synchronized to millisecond-level accuracy so that cross-container timestamps are correct when computing cross-module latencies (\eg $t_e - t_s$ in \eqnref{eqn:design_latency_done}) for dropping decisions.
}

\response{B/D}{Expand \secref{sec:eval_methodology} with more implementation details.}

\label{page:workload_app}
\parab{Workload.} Following prior works~\cite{shen2019nexus, hu2021scrooge, crankshaw2020inferline, hsieh2018focus}, we build three real-world pipelines: 
(1) \texttt{tm} (traffic monitoring), which uses three models (object detection, face recognition, and text recognition) to monitor vehicle and pedestrian information. 
(2) \texttt{lv} (live video analysis), which analyzes live video using five models (person detection, face recognition, expression recognition, eye tracking, and pose recognition). 
(3) \texttt{ga} (game analysis), which analyzes game streaming using five models (object detection, kill count detection, alive player recognition, health value recognition, and icon recognition). 
\revise{We also provide APIs for new pipeline registration.}{We also build a DAG-style pipeline based on \texttt{lv}: (4) \texttt{da} (DAG-style live video analysis), where requests from person detection module are simultaneously forwarded to the pose recognition and face recognition modules, and their outputs are subsequently merged in the expression recognition module.}

\response{A/B/D/E}{Build a DAG-style pipeline and evaluated in \secref{sec:eval_overall}.}

\label{page:workload_trace}
\revise{
    Following existing systems [cite: crankshaw2020inferline, hu2021scrooge, shen2019nexus], we set the latency objectives for the above three applications as 400ms, 500ms, and 600ms, respectively, which gives the serving system flexibility to enable dynamic batching for resource efficiency. Input video streams are collected from public datasets [cite] and video streaming platforms [cite]. We utilize two real-world traces as input rates: the Wikipedia access trace [cite: urdaneta2009wikipedia] and the Twitter access trace [cite: khaleq2018cloud]. Both traces exhibit periodic and bursty characteristics typical of DNN inference workloads as shown in \figref{fig:eval_realtime_goodput2_norm}a and \figref{fig:eval_realtime_goodput2_norm}e.
}{%
    Following prior work~\cite{crankshaw2020inferline, hu2021scrooge, shen2019nexus}, we set latency SLOs for the pipelines to 400ms, 500ms, 600ms, and 420ms, respectively, enabling flexibility for dynamic batching and resource efficiency. We also conduct a sensitivity study in \secref{sec:eval_sensitivity} to evaluate how \sysname performs under tighter and looser SLOs. Input video streams are collected from public datasets~\cite{barman2018gamingvideoset, oliveira2021vehicle} and video streaming platforms~\cite{youtube, twitch}. As shown on the left of \figref{fig:eval_realtime_goodput2_norm}, we replay three representative real-world traces as request rates: the Wikipedia access trace~\cite{urdaneta2009wikipedia}, the Twitter access trace~\cite{khaleq2018cloud}, and the Azure Function trace~\cite{shahrad2020serverless}. These traces capture typical periodic and bursty patterns of DNN inference workloads and are widely used in evaluating inference serving systems~\cite{gunasekaran2022cocktail, strati2024orion, yang2022infless}.
}

\response{E}{Highlight that we provide an SLO sensitivity analysis in \secref{sec:eval_sensitivity}.}
\response{E}{Evaluate \sysname on Azure Function trace.}


\begin{table}[t]
    \footnotesize
    \centering
    \setlength{\tabcolsep}{4pt}
    \caption{\revise{}{Ablation baselines considered in \secref{sec:eval_ablation}.}}
    \label{tab:ablation_baselines}
    \revise{}{
        \begin{tabular}{l  l  l}
        \toprule
        \textbf{Ablation}        & \textbf{Source}                              & \textbf{Key characteristic} \\
        \midrule
        \sysname-back            & \cite{hu2021scrooge, gujarati2020serving}    & Considers preceding modules only \\
        \sysname-sf              & \cite{kim2023dream}                          & Ignores $Q,W$ of subsequent modules \\
        \sysname-oc              & \cite{zhou2018overload}                      & Overload control based on $Q$ \\
        \sysname-split           & \cite{crankshaw2017clipper}                  & Fixed per-module SLO split \\
        \sysname-WCL             & --                                           & Split latency budget dynamically \\
        \sysname-lower           & --                                           & Assumes batch wait as $0$ \\
        \sysname-upper           & --                                           & Assumes batch wait as $\Sigma d_i$ \\
        \sysname-FCFS            & \cite{shen2019nexus}                         & Drops by arrival order \\
        \sysname-HBF             & --                                           & High-Budget-First only \\
        \sysname-LBF             & \cite{zhang2023shepherd}                     & Low-Budget-First only \\
        \bottomrule
        \end{tabular}
    }
\end{table}

\label{page:baselines}
\parab{Baseline.} We use Nexus~\cite{shen2019nexus} and Clipper~\cite{crankshaw2017clipper} as primary baselines because they are open-source and implement dropping policies.
\revise{
    Besides Nexus and Clipper, some non-open-source systems explicitly describe their request dropping policies [cite: kim2023dream, gujarati2020serving, hu2021scrooge]. We manually implemented their dropping policies and compared them with \sysname in \secref{sec:eval_ablation} as ablation baselines. Nexus employs a reactive dropping policy that scans the request queue in arrival order using a sliding window equal to the current batch size. It stops at the first position where all requests in the window have sufficient remaining latency to complete the inference of the current module and drops all previous requests. Clipper, originally designed for single-module applications, drops requests only if their latency exceeds the latency objective before inference. Similar to [cite: shen2019nexus] we extended Clipper to Clipper++ for inference pipelines by proportionally dividing the end-to-end SLO among modules based on their execution durations, \ie ${SLO}_k = SLO * d_k / \sum_{i=1}^{N}d_i$, and making drop decisions based on these per-module budgets. We also included a naive baseline, Naive, which has no dropping policy, to highlight the importance of request dropping.
}{%
    Nexus adopts a reactive policy that scans the queue in arrival order with a sliding window equal to the batch size, stopping at the first position where all requests in the window can meet the current module’s latency budget and dropping all earlier ones. Clipper, designed for single-module applications, drops requests only if they already exceed the latency objective before inference. Following~\cite{shen2019nexus}, we extend Clipper to Clipper++ for pipelines by proportionally dividing the end-to-end SLO across modules, \ie ${SLO}_k = SLO * d_k / \sum_{i=1}^{N} d_i$, and making per-module drop decisions accordingly. We also include a naive baseline, which applies no dropping policy.

    Several recent serving systems also incorporate request dropping~\cite{kim2023dream, zhang2023shepherd, hu2021scrooge, gujarati2020serving, zhou2018overload}. To isolate each design in \sysname, we construct ablation baselines by disabling or replacing components with those from these systems. \tabref{tab:ablation_baselines} summarizes these baselines, with detailed explanations in \secref{sec:eval_ablation}.
}

\response{C/D/E}{Compare with more strong and recent baselines (\tabref{tab:ablation_baselines}).}

\parab{Metrics.} We focused on three metrics: (1) \textit{Goodput}: the number of requests completed within the latency objective per unit time, indicating quality of service. We focus on periods of the workload that require request dropping and provide the goodput during these periods. (2) \textit{Drop rate}: the ratio of dropped requests to total requests. Requests that have completed inference but violate the SLO are also considered dropped requests. (3) \textit{Invalid rate}: the ratio of invalid computation to total computation, defined as the ratio of GPU time consumed by dropped requests to the total GPU time consumed by all requests. An optimal inference system should enhance goodput while minimizing drop and invalid rates.


\subsection{Comparison Results} \label{sec:eval_overall}

\label{page:overall_results}
\begin{figure}[t]
    \centering
    \includegraphics[width=0.47\textwidth]{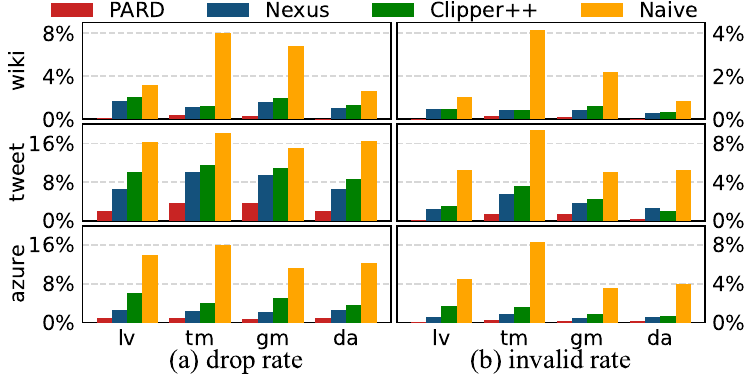}
    \caption{\revise{}{Average drop rate and invalid rate of \sysname and baseline systems under $12$ workloads.}} \narrow{}
    \label{fig:eval_overall}
    \Description{}
\end{figure}

\parab{\revise{}{Overall Results.}}
\revise{
    \figref{fig:eval_realtime_goodput2_norm} shows the normalized goodput of \sysname and the three alternatives, where \sysname consistently achieves a higher goodput than all alternatives during the whole period. Specifically, as shown in \figref{fig:eval_overall}, \sysname drops an average of only $0.07\%$-$3.6\%$ of requests under various workloads, which reduces the drop rate and wasted computation resources by $1.6\times$-$34\times$ and $1.5\times$-$61\times$ than Nexus and Clipper++. Therefore, as shown in \figref{fig:eval_realtime_goodput2_norm}, \sysname increases the goodput by $21\%$-$176\%$ compared to Nexus and Clipper++, indicating the effectiveness of its proactive dropping policy. Besides, the naive baseline has the worst goodput under all six workloads with an average drop rate and invalid rate $4.9\times$-$27.2\times$ and $12.7\times$-$128.8\times$ of \sysname's, indicating the importance of request dropping for goodput enhancement.
}{%
    \figref{fig:eval_realtime_goodput2_norm} presents the normalized goodput of \sysname compared with three alternatives, where \sysname maintains higher goodput across nearly the entire period. Specifically, as shown in \figref{fig:eval_overall}, \sysname drops an average of only $0.12\%$-$3.6\%$ of requests under various workloads, which reduces the drop rate and wasted computation resources by $1.6\times$-$16.7\times$ and $1.5\times$-$61.9\times$ than Nexus and Clipper++. Therefore, as shown in \figref{fig:eval_realtime_goodput2_norm}, \sysname increases the goodput by $16\%$-$176\%$ compared to Nexus and Clipper++, indicating the effectiveness of its proactive dropping policy. Besides, the naive baseline has the worst goodput under all six workloads, with average drop and invalid rates up to $35\times$ and $129\times$ those of \sysname, highlighting the importance of request dropping for goodput enhancement.
}

\begin{figure}[t]
    \centering
    \includegraphics[width=\linewidth]{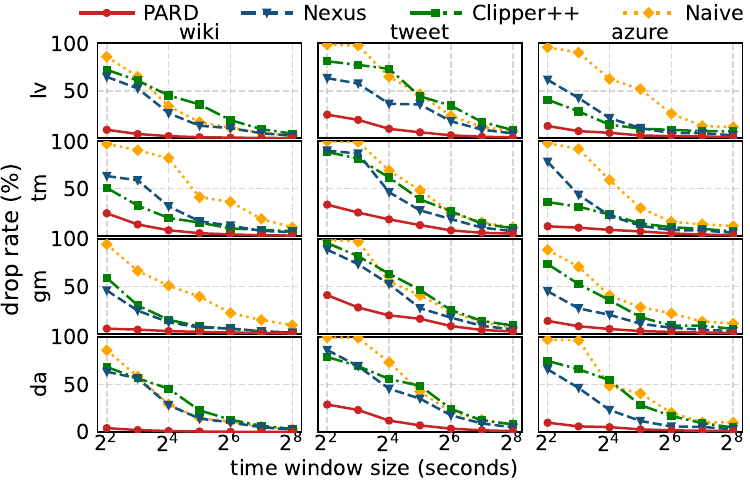}
    \caption{\revise{}{The maximum average drop rate over the entire runtime in different time window sizes.}}\narrow{}
    \label{fig:eval_droprate_timescale}
    \Description{}
\end{figure}

\begin{figure*}[t]
    \centering
    \subfloat{
        \includegraphics[width=0.216\linewidth]{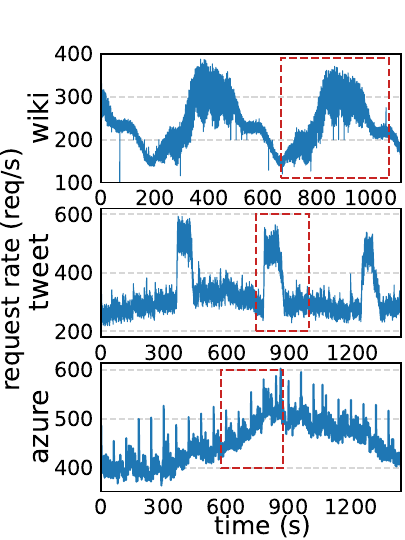}
    }
    \subfloat{
        \includegraphics[width=0.768\linewidth]{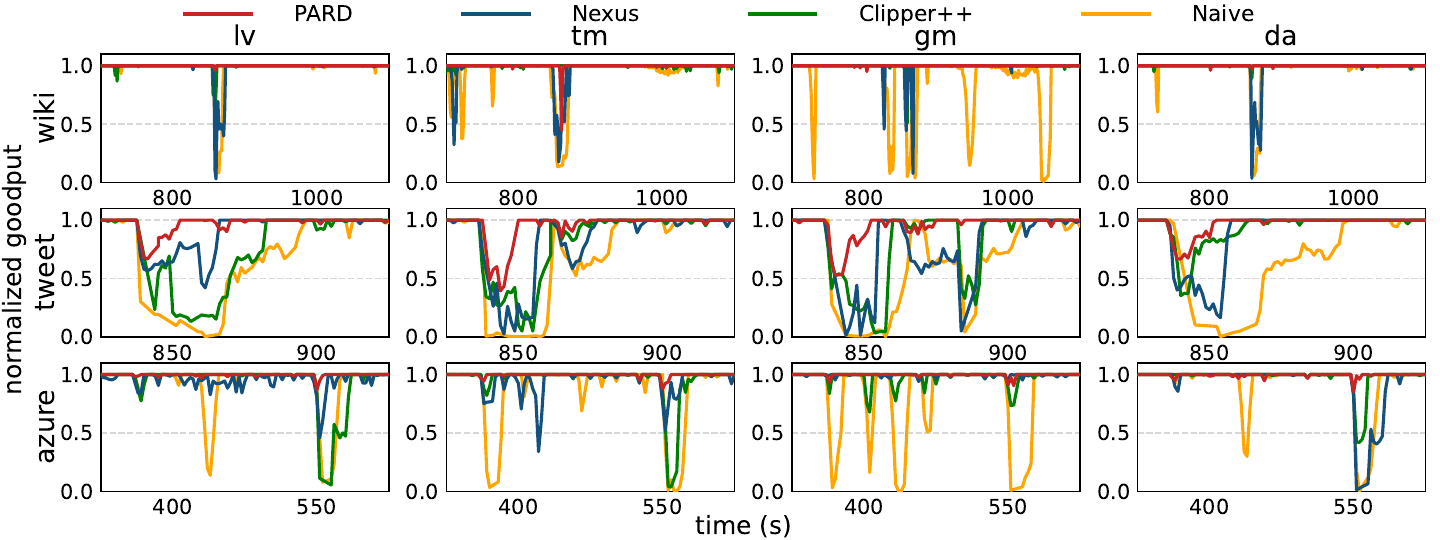}
    }
    \caption{\revise{}{Left: Three real-world traces~\cite{urdaneta2009wikipedia, khaleq2018cloud, shahrad2020serverless}. Right: The normalized real-time goodput of \sysname and baseline systems across $12$ workloads, corresponding to the red-boxed regions in each trace.}}
    \label{fig:eval_realtime_goodput2_norm}
    \Description{}
\end{figure*}


\revise{
    To understand the performance differences, we conjecture that \sysname's advantage arises from two key factors as follows. We also provide a detailed ablation study in \secref{sec:eval_ablation} to quantify the impact of each design choice on goodput enhancement.
    
    First, alternative systems can \textit{not} properly decide when to drop requests for each module in the inference pipeline. Nexus adopts a reactive drop policy, making dropping decisions without considering the latency budget for subsequent modules, leading to $57.1\%$ to $97.2\%$ of dropped requests concentrated in the latter half of the pipeline. While Clipper++ can identify requests with insufficient latency budget earlier than Nexus by splitting the end-to-end latency objective into per-module latency budgets, leading to $39.6\%$-$78.5\%$ of dropped requests concentrated in the latter half of the pipeline. However, as shown in \figref{fig:eval_overall}, it suffers from a higher invalid rate, surpassing both \sysname and Nexus, due to latency fragmentation caused by the latency splitting (\secref{sec:eval_ablation}). In contrast, \sysname implements a proactive drop policy with bi-directional request dropping, enabling timely dropping decisions based on end-to-end latency estimation. \sysname also derives a sweet spot for the unpredictable $w_k$ to balance between mis-dropped and mis-kept requests. By doing so, \sysname achieves lower invalid rates than Nexus and Clipper++, and reduces drop rates due to smaller request backpressure, ultimately leading to higher goodput. 
}{%
    To understand the performance gap, we attribute the advantage of \sysname to two factors, further validated by the ablation study in \secref{sec:eval_ablation}. First, existing systems fail to decide \textit{when} to drop requests across modules. Nexus makes reactive dropping decisions without considering latency budget for subsequent modules, causing $51\%$–$97\%$ of drops to cluster in the latter half of the pipeline. Clipper++ identifies requests with insufficient latency budget earlier by splitting the latency SLO into per-module budgets. However, $31\%$–$89\%$ of drops still occur in later modules, and the system suffers higher invalid rates than both Nexus and \sysname since splitting restricts latency budget flexibility (\secref{sec:eval_ablation}). In contrast, \sysname's proactive dropping enables early decisions based on end-to-end latency estimation. Thus \sysname achieves lower invalid rates and drop rates, leading to higher goodput.
}

\response{A/B/D/E}{Extend the analysis with results on \texttt{azure} trace and \texttt{da} application.}

\revise{
    Second, alternative systems can \textit{not} select the proper set of requests to drop for each module. Both Nexus and Clipper++ make dropping decisions based on the request arrival order, regardless of each request's remaining latency budget and workload intensity. As discussed in \secref{sec:motivation_analysis}, this could excessively consume the latency budget during workload burst and bring unnecessary queueing delay to requests with lower remaining latency budgets, leading to a transient drop rate of $45.9\%$-$95.8\%$ as shown in \figref{fig:eval_droprate_timescale}. In contrast, \sysname reorders requests dynamically using adaptive priority to selectively drop certain requests, enhancing the overall goodput. By doing so, \sysname reduces the drop rate by $44.5\%$-$96.9\%$ at any time scale compared to Nexus and Clipper++ (\figref{fig:eval_droprate_timescale}) and achieves higher goodput.
}{%
    Second, existing systems fail to select \textit{which} requests to drop properly. Both Nexus and Clipper++ drop requests purely in arrival order, ignoring remaining latency budgets and workload intensity. This design wastes latency budgets during bursts and adds unnecessary queueing delays for requests with tighter deadlines, yielding transient drop rates up to $90\%$ and $96\%$ (\figref{fig:eval_droprate_timescale}). Instead, \sysname dynamically reorders requests using adaptive priority for selective dropping decisions, cutting transient drop rates by $41\%$–$98\%$ across all timescales and consistently improving goodput (\figref{fig:eval_droprate_timescale}).

}

\label{page:DAG_results}
\parab{\revise{}{Applicable to DAG workloads.}}
\revise{}{
    In DAG-style pipelines, \sysname estimates the maximum end-to-end latency across all branches and drops requests accordingly. As shown in \figref{fig:eval_overall}a, this yields $2.1\times$–$12.6\times$ and $3.1\times$–$16.7\times$ lower drop rates than Nexus and Clipper++ under application \texttt{da}. Unlike \texttt{lv}, \texttt{da} executes pose and face recognition modules in parallel; dropping in one module invalidates computation in the other one for \texttt{da}. This raises \sysname’s invalid rate to $1.21\times$–$1.36\times$ that of \texttt{lv}, though still $3.2\times$–$15.9\times$ lower than baselines.
    

    Recent DAG pipelines further complicate request dropping with request-specific dynamic paths~\cite{strati2024orion, lu2024smiless, bhasi2021kraken}, where the chosen branch depends on intermediate results. This variability amplifies latency uncertainty and reduces the accuracy of \sysname’s dropping decisions. To evaluate this, we adapt \texttt{da} so each request probabilistically takes either the pose or face branch. In this setting, \sysname’s drop rate rises by $0.05\times$, $0.21\times$, and $0.10\times$ across three traces due to mis-estimation. Request-path prediction techniques~\cite{daw2020xanadu, stojkovic2023specfaas} could be incorporated into \sysname to yield more accurate latency estimation, which we leave for future work.

}

\response{A/B/D/E}{Extend the analysis on both fixed path and dynamic path DAG workloads.}


\subsection{Ablation Study} \label{sec:eval_ablation}


In this section, we disable each of \sysname's features to demonstrate its contribution to guaranteeing goodput\footnote{Note that we use \texttt{lv-tweet} workload for ablation study and sensitivity analysis. Other workloads show similar results and are omitted for brevity.}.

\response{none}{Remove previous Figure 12a (Normalized goodput distribution) for page limit.}

\label{page:ablation}


\parab{How important is \revise{bi-directional}{proactive latency estimation}?}
\revise{
     \sysname outperforms alternative systems due to its bi-directional request dropping, which estimates end-to-end latency for timely decisions (\secref{sec:eval_overall}). To verify that, we compare \sysname with (1) \sysname-back, considering only preceding and current modules (\ie $\mathbb{L}_{sub} = 0$), similar to Nexus [cite], Scrooge [cite] and Clockwork [cite], (2) \sysname-sf, including execution duration for subsequent modules (\ie $\mathbb{L}_{sub} = \sum_{i=k+1}^{N}d_i$), similar to DREAM [cite].
}{%
    \sysname outperforms alternatives by making timely drop decisions using end-to-end latency estimates derived from \textit{bi-directional runtime information} (\secref{sec:eval_overall}). To verify that, we compare against (1) \sysname-back, which considers only preceding and current modules (\ie $\mathbb{L}_{sub} = 0$), similar to Clockwork~\cite{gujarati2020serving}, Nexus~\cite{shen2019nexus}, and Scrooge~\cite{hu2021scrooge}; (2) \sysname-sf, which accounts for execution durations of subsequent modules (\ie $\mathbb{L}_{sub} = \sum_{i=k+1}^{N}d_i$), similar to DREAM~\cite{kim2023dream}; and (3) \sysname-oc, which adopts DAGOR's overload control strategy~\cite{zhou2018overload}. In \sysname-oc, requests are dropped when the average queueing delay of a module exceeds a threshold $T$; the module then notifies preceding modules and admits requests at a rate of $(1-\alpha)\times\text{input\_rate}$ simultaneously\footnote{We tune $T$ and $\alpha$ for each traces, and obtain the best performance with $\alpha=0.4$, $T=20$ms for the \texttt{wiki}, and $T=25$ms for \texttt{tweet} and \texttt{azure}.}.
}

\begin{figure}[t]
    \centering
    \subfloat[Drop rate and invalid rate]{
        \includegraphics[width=0.23\textwidth]{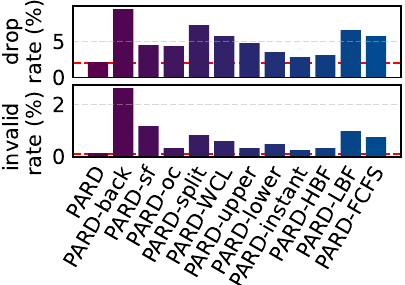}
        \label{fig:ablation_all}
    }
    \subfloat[Drops at each module]{
        \includegraphics[width=0.23\textwidth]{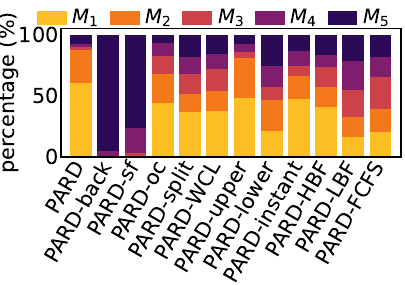}
        \label{fig:ablation_A_dropat}
    }
    \caption{\revise{}{Comparison results of \sysname and its alternatives: (a) Average drop rate and invalid rate. (b) Percentage of dropped requests at each module.}}\narrow{}
    \Description{}
\end{figure}

\begin{figure*}[t]
    \centering
    \subfloat[Consumed latency budget]{
        \includegraphics[width=0.23\textwidth]{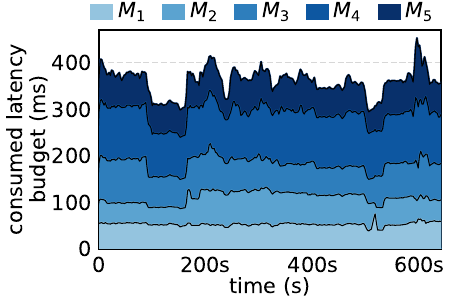}
        \label{fig:ablation_A_consumed}
    }
    \subfloat[Latency distribution]{
        \includegraphics[width=0.23\textwidth]{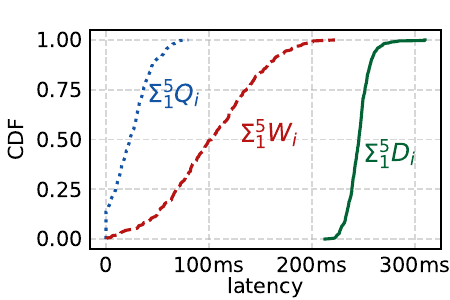}
        \label{fig:ablation_B_batch_wait}
    }
    \subfloat[Queueing delay]{
        \includegraphics[width=0.23\textwidth]{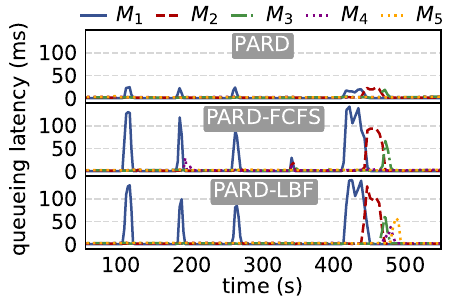}
        \label{fig:ablation_C_queueing}
    }
    \subfloat[Remaining latency budget]{
        \includegraphics[width=0.23\textwidth]{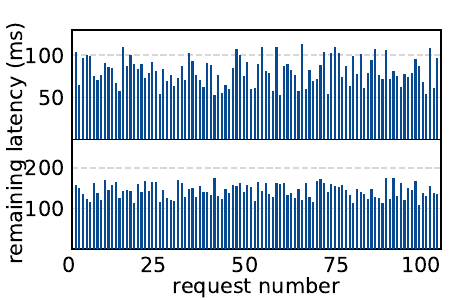}
        \label{fig:ablation_C_remaining}
    }
    \caption{(a) Consumed latency budget of $100k$ SLO-compliant requests at each module. (b) CDF of end-to-end queueing delay $\sum_{i=k+1}^{N}Q_i$, batch wait time $\sum_{i=k+1}^{N}W_i$, and inference duration $\sum_{i=k+1}^{N}D_i$. (c) Queueing delay of each module in the pipeline during workload burst. (d) The remaining latency budget of $100$ consecutive requests at $M_2$ (lower) and $M_3$ (upper).}\narrow{}
    \label{fig:ablation_data}
    \Description{}
\end{figure*}

\revise{
    As shown in \figref{fig:ablation_all}, \sysname-back and \sysname-sf exhibit drop rates $3.6\times$ and $1.2\times$ higher than \sysname's, with invalid rates $24\times$ and $9.8\times$ higher, respectively. \sysname-back, which neglects the latency budget for subsequent modules, has an average drop rate of $9.5\%$ and can spike up to $98\%$, concentrating $95\%$ of drops in the last module and resulting in the highest invalid rate (\figref{fig:ablation_A_dropat}). \sysname-sf performs better, with a drop rate of $4.5\%$ and $76\%$ of drops in the last module, but still suffers from the drop-too-late issue due to ignoring queueing delay and batch wait time in subsequent modules (\ie $\sum_{i=k+1}^{N}Q_i$ and $\sum_{i=k+1}^{N}W_i$ in $\mathbb{L}_{sub}$), highlighting their significance for enhancing goodput.
}{%
    

    As shown in \figref{fig:ablation_all}, \sysname-back, \sysname-sf, and \sysname-oc yield drop rates $1.1\times$-$3.6\times$ higher than \sysname, with invalid rates $2.1\times$-$24\times$ higher. \sysname-back ignores downstream budgets, causing an average drop rate of $9.5\%$, and $95\%$ of drops concentrated in the last module, producing the highest invalid rate (\figref{fig:ablation_A_dropat}). \sysname-sf improves to a $4.5\%$ drop rate with $76\%$ of drops in the last module, but still suffers late drops by neglecting subsequent queueing and batch wait ($\sum_{i=k+1}^{N} Q_i$ and $\sum_{i=k+1}^{N} W_i$), highlighting their importance. \sysname-oc achieves fewer late drops by jointly throttling admission across modules, with only $7.2\%$ requests being dropped in the last module. However, its microservice-oriented design overlooks the large latency uncertainty from batching. This coarse-grained design still yields drop and invalid rates $2.1\times$ and $3.0\times$ higher than \sysname, highlighting the need for proactive latency estimation in DNN pipelines.
    
}

\revise{
    In contrast, \sysname, by accurately estimating $\mathbb{L}_{sub}$, concentrates $87\%$ of drops in the first two modules, reducing invalid rates by $24\times$ and $9.8\times$ than \sysname-back and \sysname-sf. The reduced computation benefits other requests, reducing the average queueing delay by $12\%$ and $5.9\%$ over \sysname-back and \sysname-sf. Consequently, \sysname's bi-directional policy reduces drop rates by up to $78\%$ and $54\%$.
}{%
    In contrast, \sysname, by accurately estimating $\mathbb{L}_{sub}$, concentrates $87\%$ of drops in the first two modules, reducing invalid rates by $2.1\times$ to $24\times$ than three baselines. The reduced computation benefits other requests, reducing the average queueing delay by $4.7\%$ to $12\%$. Consequently, \sysname's proactive policy with bi-directional runtime information reduces drop rates by up to $52\%$ to $78\%$.
}

\response{D/E}{Add an stronger ablation baseline based on overload control approach.}


\parab{Why not split the latency objective?} 
\revise{
    \sysname compares the request's estimated end-to-end latency $\mathbb{L}$ against the latency objective for request dropping. Alternatively, the latency objective can be split into per-module budgets and then make dropping decisions by comparing requests' latency in each module to its budget. We argue that this latency-splitting approach can \textit{not} drop requests properly. To verify that, we compare \sysname with \sysname-split, which splits the SLO into per-module budgets, similar to Clipper++ [cite: crankshaw2017clipper, shen2019nexus].
}{%
    \sysname drops requests by comparing estimated end-to-end latency $\mathbb{L}$ with the overall SLO. An alternative is to split the SLO into per-module budgets and drop requests when their latency exceeds a module’s budget. We argue this approach cannot drop requests properly. To validate, we compare: (1) \sysname-split, which splits the SLO into fixed per-module budgets, similar to Clipper++; and (2) \sysname-WCL, which dynamically allocates budgets based on modules' runtime worst-case latency (WCL), including queueing, batch wait, and execution.
    
}

\revise{
    As shown in \figref{fig:ablation_all}, \sysname-split's drop rate and invalid rate are $4.7\times$ and $15.7\times$ higher than \sysname's. This is because the fixed per-module latency budgets cause latency budget fragmentation, where unused budgets in one module cannot be utilized by others. We measure \sysname-split's residual latency budget, \ie the difference between the end-to-end latency objective and the actual request latency, which is $16.7\%$-$27.9\%$ larger than \sysname's, indicating \sysname-split's inefficiency in utilizing the latency budget. Although \sysname-split prevents request drops from being concentrated in later modules by limiting the maximum latency consumption of each module (\figref{fig:ablation_A_dropat}), its fragmentation leads to excessive drops across \textit{all} modules, resulting in the highest drop rate and a $42\%$ lower goodput than \sysname (\figref{fig:ablation_goodput}).

    Moreover, we argue that any policy assigning fixed latency budgets to each module and making independent dropping decisions cannot achieve high goodput. \figref{fig:ablation_A_consumed} shows the actual consumed latency for $100k$ SLO-compliant requests, which varies over time due to different batch wait times and queueing delays in each module. Thus, no fixed assigning policy can align with the latency consumption of all requests, and the inference system must consider the entire pipeline as a whole rather than treating each module independently.
}{%

    As shown in \figref{fig:ablation_all}, \sysname-split incurs drop and invalid rates $2.6\times$ and $6.7\times$ higher than \sysname. Splitting prevents early modules from over-consuming budgets, concentrating $51\%$ of drops in the first two modules (\figref{fig:ablation_A_dropat}) and keeping the invalid rate low ($0.82\%$). However, splitting limits flexibility: when a module suffers queueing delays from cold starts (\eg around $200$s and $600$s in \figref{fig:ablation_A_consumed}), the system cannot reallocate budgets to mitigate the bottleneck. \sysname-WCL reduces drop and invalid rates by $30\%$ and $27\%$ relative to \sysname-split by dynamically allocating budgets according to WCL. While even with tuned allocate frequency, it still underperforms due to budget demands fluctuating rapidly across modules (\figref{fig:ablation_A_consumed}), and batching introduces large variation in per-request latency (\figref{fig:ablation_C_remaining}). As a result, its drop and invalid rates remain $2.8\times$ and $5.4\times$ higher than \sysname. These results show that neither static nor dynamic splitting can handle the latency uncertainty of batching in DNN pipelines. In contrast, by treating the pipeline as a whole and estimating end-to-end per-request latency, \sysname achieves $64\%$ and $72\%$ lower drop rates than \sysname-split and \sysname-WCL.
    

}

\response{C/D/E}{Refactor baseline \sysname-split and add stronger baseline \sysname-WCL.}

\parab{Why use a sweet spot $w_k$?}
\revise{
    Since the aggregated batch wait time $\sum_{i=k+1}^{N}W_i$ is unpredictable and highly variable, \sysname derives a sweet spot $w_k$ to approximate $\sum_{i=k+1}^{N}W_i$ for dropping decisions. To verify that, we compare \sysname with (1) \sysname-lower, which estimates $\sum_{i=k+1}^{N}W_i$ as its lower bound $0$ (\ie $\mathbb{L}_{sub} = \sum_{i=k+1}^{N}Q_i + \sum_{i=k+1}^{N}d_i$); (2) \sysname-upper, which estimates $\sum_{i=k+1}^{N}W_i$ as its upper bound $\sum_{i=k+1}^{N}d_i$ (\ie $\mathbb{L}_{sub} = \sum_{i=k+1}^{N}Q_i + 2\sum_{i=k+1}^{N}d_i$).

    As shown in \figref{fig:ablation_all}, \sysname-lower's invalid rate is $3.5\times$ higher than \sysname since \sysname-lower under-estimates $\sum_{i=k+1}^{N}W_i$ as $0$, which mis-keeps requests with insufficient latency budget to the later modules. Therefore, according to \figref{fig:ablation_A_dropat}, the percentage of dropped requests of \sysname-lower is $47\%$ lower than \sysname's in the first two modules but $3.2\times$ larger in the last three modules, resulting in a goodput $6.8\%$ lower than \sysname. Meanwhile, \sysname-upper achieves an average drop rate $1.3\times$ higher than \sysname since it over-estimates $\sum_{i=k+1}^{N}W_i$ as $\sum_{i=k+1}^{N}d_i$, leading to mis-drops of requests with sufficient latency budget. As shown in \figref{fig:ablation_all} and \figref{fig:ablation_A_dropat}, \sysname-upper drops $82\%$-$179\%$ more requests throughout each module in the inference pipeline, leading to a goodput $6.5\%$ lower than \sysname's. 
    
    \figref{fig:ablation_B_batch_wait} further shows the distribution of the elements in $\mathbb{L}_{sub}$ for $100k$ requests, where $\sum_{i=k+1}^{N}W_i$ has a much larger variance compared to $\sum_{i=k+1}^{N}Q_i$ and $\sum_{i=k+1}^{N}D_i$. Therefore, \sysname try to derive an estimated $w_k$ between $0$ and $\sum_{i=k+1}^{N}d_i$ to represent the unpredictable $\sum_{i=k+1}^{N}W_i$. Although some requests will still be mis-kept or mis-dropped, this approach allows \sysname to balance the drop rate and invalid rate with the sweet spot $w_k$, thereby enhancing the goodput. 
}{%
    Because the aggregated batch wait time $\sum_{i=k+1}^{N} W_i$ is unpredictable and highly variable, \sysname derives a sweet spot $w_k$ to estimate it for dropping decisions. To validate this design, we compare \sysname with: (1) \sysname-lower, which assumes the lower bound $\sum_{i=k+1}^{N} W_i=0$ (\ie $\mathbb{L}_{sub}=\sum_{i=k+1}^{N} Q_i + \sum_{i=k+1}^{N} d_i$); and (2) \sysname-upper, which assumes the upper bound $\sum_{i=k+1}^{N} W_i=\sum_{i=k+1}^{N} d_i$ (\ie $\mathbb{L}_{sub}=\sum_{i=k+1}^{N} Q_i + 2\sum_{i=k+1}^{N} d_i$).

    As shown in \figref{fig:ablation_all}, \sysname-lower under-estimates $\sum_{i=k+1}^{N} W_i$, mis-keeping requests that lack sufficient budgets for later modules. Its invalid rate is $3.5\times$ higher than \sysname, with $3.2\times$ more drops concentrated in the last three modules (\figref{fig:ablation_A_dropat}), lowering goodput by $6.8\%$. In contrast, \sysname-upper over-estimates $\sum_{i=k+1}^{N} W_i$, mis-dropping requests with adequate budgets. It yields a drop rate $1.3\times$ higher, leading to $6.5\%$ lower goodput. \figref{fig:ablation_B_batch_wait} further shows that $\sum_{i=k+1}^{N} W_i$ exhibits far greater variance than $\sum_{i=k+1}^{N} Q_i$ or $\sum_{i=k+1}^{N} D_i$. Hence, \sysname derives $w_k$ within $[0, \sum_{i=k+1}^{N} d_i]$ to estimate $\sum_{i=k+1}^{N} W_i$. While some requests are still mis-kept or mis-dropped, this sweet-spot estimation balances drop and invalid rates, thereby improving goodput.
}

\response{none}{Shorten the analysis for page limit.}

\begin{figure}[t]
    \centering
    \includegraphics[width=0.47\textwidth]{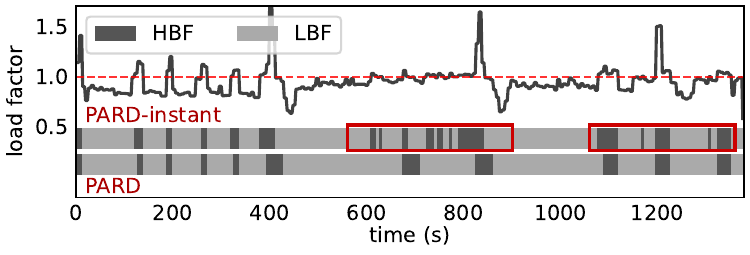}
    \caption{Load factor and prioritization mechanism transition in \sysname and \sysname-instant.}\narrow{}
    \label{fig:ablation_C_factor}
    \Description{}
    \vspace{-6pt}
\end{figure}

\parab{Why does the remaining latency budget matter?}
\revise{
    Reactive policies drop requests by arrival order, leading to suboptimal decisions. \sysname, in contrast, determines the dropping set based on the remaining latency budget and workload intensity, ensuring goodput under varying workloads. To verify that, we compare \sysname with (1) \sysname-FCFS, which drops requests in arrival order, similar to Nexus and Clipper++; (2) \sysname-HBF, which uses only HBF to reorder requests; (3) \sysname-LBF, which uses only LBF; and (4) \sysname-instant, which uses both HBF and LBF without delayed transition.
}{%
    Reactive policies drop requests in arrival order, leading to suboptimal decisions. In contrast, \sysname selects the dropping set based on remaining latency budgets and workload intensity, sustaining goodput under diverse workloads. To verify this, we compare with: (1) \sysname-FCFS, which drops by arrival order like Nexus~\cite{shen2019nexus} and Clipper++; (2) \sysname-HBF, which always applies HBF; (3) \sysname-LBF, which always applies LBF, similar to SHEPHERD~\cite{zhang2023shepherd}; and (4) \sysname-instant, which applies HBF and LBF without delayed transition.

}

\response{D/E}{Add reference for the baselines that share the same policy with recent systems.}

\begin{figure*}[t]
    \centering
    \subfloat[Stress-testing]{
        \includegraphics[width=0.2836\textwidth]{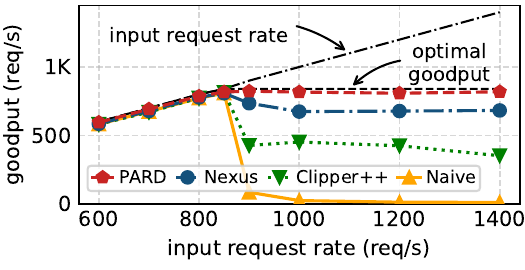}
        \label{fig:eval_sensitive_qps}
    }
    \subfloat[SLO sensitivity]{
        \includegraphics[width=0.2188\textwidth]{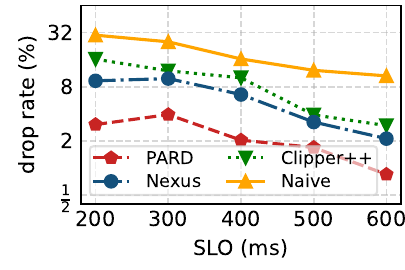}
        \label{fig:eval_sensitive_slo}
    }
    \subfloat[\revise{}{Sensitivity of $\lambda$}]{
        \includegraphics[width=0.2188\textwidth]{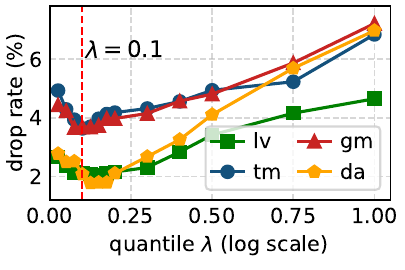}
        \label{fig:eval_sensitive_lambda}
    }
    \subfloat[\revise{}{Sensitivity of window size}]{
        \includegraphics[width=0.2188\textwidth]{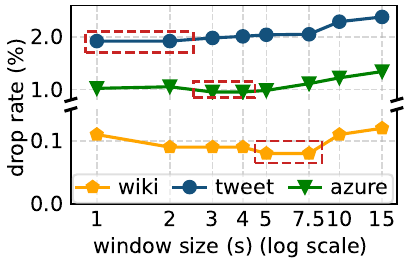}
        \label{fig:eval_sensitive_window}
    }
    \vspace{-3pt}
    \caption[]{
        Experimental results for sensitivity analysis: (a) Average goodput across different request rates. (b) Average drop rate under various SLO settings. \revise{}{(c) Drop rate at different quantile $\lambda$. (d) Drop rate at different moving window sizes\footnotemark.}
    }\narrow{}
    \label{fig:eval_sensitive}
    \Description{}
\end{figure*}

\revise{
    As shown in \figref{fig:ablation_all}, the drop rates for \sysname-FCFS, \sysname-LBF, and \sysname-HBF are $1.8\times$, $2.2\times$, and $0.5\times$ higher than \sysname's due to several factors. (1) During workload bursts, \sysname-FCFS always prioritizes the earliest arrived requests, over-consuming subsequent requests' latency budgets and causing accumulation (\figref{fig:ablation_C_queueing}). This increases queueing delay by $34\%$ and reduces goodput by $24\%$ compared to \sysname. Similarly, \sysname-LBF suffers from request accumulation, resulting in a goodput $29\%$ lower than \sysname's. (2) During steady workloads, \sysname-FCFS cannot make proper dropping decisions due to latency uncertainty. \figref{fig:ablation_C_remaining} shows highly variable, time-independent remaining latency budgets for $100$ requests in $M_2$ and $M_3$. Both \sysname-FCFS and \sysname-HBF ignore this variability, leading to goodput reductions of $24\%$ and $6\%$, respectively, compared to \sysname (\figref{fig:ablation_goodput}).

    Though \sysname-instant achieves the lowest drop rate among the alternatives, it still drops $25\%$ more requests than \sysname due to frequent transition between HBF and LBF during workload fluctuations, as highlighted by the red rectangles in \figref{fig:ablation_C_factor}. Conversely, \sysname provides a smooth transition between HBF and LBF, resulting in the highest goodput.
}{%
    As shown in \figref{fig:ablation_all}, the drop rates for \sysname-FCFS, \sysname-LBF, and \sysname-HBF are $1.8\times$, $2.2\times$, and $0.5\times$ higher than \sysname's due to several factors. (1) During workload bursts, \sysname-FCFS always prioritizes the earliest arrived requests, over-consuming subsequent requests' latency budgets and causing accumulation (\figref{fig:ablation_C_queueing}). This increases queueing delay by $34\%$ and reduces goodput by $24\%$ compared to \sysname. Similarly, \sysname-LBF suffers from request accumulation, resulting in a goodput $29\%$ lower than \sysname's. (2) During steady workloads, \sysname-FCFS fails to make proper dropping decisions due to latency uncertainty. \figref{fig:ablation_C_remaining} shows highly variable, time-independent remaining latency budgets for $100$ requests in $M_2$ and $M_3$. Ignoring this variability, \sysname-FCFS and \sysname-HBF achieve $24\%$ and $6\%$ lower goodput than \sysname. Although \sysname-instant achieves the lowest drop rate among the baselines, it still drops $25\%$ more requests than \sysname due to frequent transition between HBF and LBF during fluctuations (\figref{fig:ablation_C_factor}). Conversely, \sysname’s delayed transition provides smooth switching and achieves the highest goodput.
}

\response{none}{Shorten the analysis for page limit.}

\subsection{Sensitivity and Overhead Analysis}\label{sec:eval_sensitivity}

\parab{Stress-testing.}
\revise{
    \sysname makes timely and precise dropping decisions based on bi-directional runtime information and workload intensity. To evaluate its performance under heavy workloads, we generate synthetic traces to stress-test the system. Specifically, we fix the total resource allocation of \sysname and the three baseline systems and gradually increase the request rate until the goodput tends to stabilize.

    As shown in \figref{fig:eval_sensitive_qps}, as the request rate increases, the baseline Naive gradually fails to meet the SLO since it cannot eliminate accumulated requests. Besides, Nexus and Clipper++ cannot handle the differences in remaining latency budgets across requests, relying solely on unidirectional information for dropping decisions. Therefore, when the request rate exceeds the processing capacity of the fixed allocated resources, \sysname achieves 11.9\%-132.9\% higher goodput compared to Nexus and Clipper++. Besides, as shown in \figref{fig:eval_sensitive_qps}, the gap between \sysname's goodput and the optimal goodput (\revise{\ie the maximum achievable goodput}{\ie the minimum of the request rate and the system throughput}) is $3.4\times-23.4\times$ smaller than that of Nexus and Clipper++.
}{%
    \sysname makes timely, precise dropping decisions using bi-directional runtime information and workload intensity. To evaluate its behavior under heavy load, we fix the instance number of \sysname and three baselines, and increase the request rate until goodput stabilizes. As shown in \figref{fig:eval_sensitive_qps}, Naive quickly fails to meet the SLO because it cannot clear accumulated requests. Nexus and Clipper++ also degrade since they rely on unidirectional information and over-consume latency budgets. When request rates exceed the testbed's capability, \sysname achieves $11.9\%$–$132.9\%$ higher goodput and the gap between \sysname’s goodput and the optimal goodput (\ie the minimum of request rate and system throughput) is $3.4\times$–$23.4\times$ smaller than that of baselines.
}


\parab{Sensitivity to SLO.}
\revise{
    We set each application's SLO following existing systems~\cite{shen2019nexus, hu2021scrooge, crankshaw2020inferline}. We evaluate the sensitivity of \sysname's advantage over existing systems by varying latency SLOs from 200ms to 600ms. To meet varying SLO constraints, \sysname, alongside existing systems, adjust the expected batch sizes of each inference pipeline module. As shown in \figref{fig:eval_sensitive_slo}, \sysname maintains average drop rates between $0.85\%$ and $3.04\%$ under different SLOs, whereas Nexus and Clipper++ exhibit average drop rates that are $1.9\times-5.3\times$ those of \sysname's, showing the robustness of \sysname's advantage over existing systems under various SLOs.
}{%
    We set each application’s SLO following prior works~\cite{shen2019nexus, hu2021scrooge, crankshaw2020inferline} and evaluate sensitivity by varying SLOs from 200ms to 600ms. To meet different constraints, all systems adjust expected batch sizes for each module. As shown in \figref{fig:eval_sensitive_slo}, \sysname sustains lowest drop rates ($0.85\%$–$3.04\%$) across SLOs, which is $1.9\times$–$5.3\times$ lower than baselines, demonstrating the robustness of \sysname’s design.
}

\response{none}{Shorten the analysis for page limit.}


\parab{Sensitivity analysis of $\lambda$.}
\revise{
    To evaluate the sensitivity of quantile $\lambda$ that determines $w_k$, we vary $\lambda$ from 0.025 to 0.2 to observe its impact on the drop rate across six workloads. As shown in \figref{fig:eval_sensitive_lambda1} and \figref{fig:eval_sensitive_lambda1}, although optimal $\lambda$ is not always 0.1, it always falls within the range of 0.075 to 0.15, and the drop rates do not exhibit significant variation. Consequently, we set $\lambda$ to $0.1$ by default and support slight adjustments according to actual performance.
}{%
    We evaluate the sensitivity of quantile $\lambda$, which determines $w_k$. As shown in \figref{fig:eval_sensitive_lambda}, the optimal $\lambda$ is not always $0.1$ but consistently lies between 0.075 and 0.15, with drop rates showing little variation within this range. Therefore, we set $\lambda=0.1$ by default.
    
}

\response{none}{Extend $\lambda$ sensitivity to a wider range.}

\label{page:sensitivity_window}
\parab{\revise{}{Sensitivity to window size.}}
\revise{}{
    \sysname applies a $5$s sliding window to smooth recent queueing delays. As in \figref{fig:eval_sensitive_window}, the optimal window is trace-dependent: bursty traces such as \texttt{tweet} (coefficient of variation, $\text{CV}\approx1.0$) and \texttt{azure} ($\text{CV}\approx1.3$) favor shorter windows due to rapidly varying queueing delays, while the stable \texttt{wiki} trace ($\text{CV}\approx0.47$) benefits from longer ones. The drop-rate gap between each trace’s optimum and the $5$s default is only $3.2\%$–$6.3\%$, confirming that the default is sufficient for our main results. For new workloads, a practical guideline is: $5$–$7$s for stable traces ($\text{CV}<0.5$), $3$–$5$s for moderately bursty traces ($0.5\le\text{CV}<1.0$), and $1$–$3$s for highly bursty traces ($\text{CV}\ge1.0$).
    
}

\response{A}{Add sensitivity analysis to the sliding window size.}

\label{page:overhead}
\parab{Overheads.}
\revise{
    \sysname's main overhead stems from batch wait distribution updates in bi-directional dropping and the DEPQ in adaptive prioritization. The runtime complexity of the batch wait distribution updates is approximately $O(MN)$, where $M$ defaults to 10k and $N$ is the number of modules in the inference pipeline. This function runs on a separate thread and is asynchronous to request latency estimation, meaning it does not introduce additional latency to requests. Besides, the complexity of the DEPQ's \texttt{put()} and \texttt{get()} operations is $O(\log n)$, where $n$ is the average length of the DEPQ. Experiments show that these operations of DEPQ increase request latency by less than 0.16\%.
}{%
    \sysname introduces three sources of overhead:
    (1) \textit{Batch-wait distribution updates}, executed asynchronously on a separate thread with complexity $O(MN)$ ($M=10{,}000$ samples over $N$ modules), adding no extra request latency.
    (2) \textit{State synchronization}, which exchanges compact module states (queueing delay, batch size, throughput, drop rate, and batch-wait distribution) once per second in a separate thread; this costs $<3.2$,Kbps per worker, negligible compared to the 240–640,Mbps data plane (image size $\times$ request rate).
    (3) \textit{Request reordering in DEPQ}, with \texttt{put()} and \texttt{get()} operations of $O(\log n)$ for average queue length $n$. Experiments show these add less than $0.16\%$ request latency.
    
}
\response{A/C}{Extend overhead analysis with state synchronization.}



\footnotetext{\revise{}{We analyze $\lambda$ sensitivity on \texttt{tweet} trace and window size sensitivity on \texttt{lv} application, omitting other workloads as they exhibit similar trends.}}

%% file: content/6_related.tex
\section{Related Work} \label{sec:related}

\label{page:related}
\parab{Request Dropping.}
\revise{
    Numerous inference systems adopt request dropping to ensure service quality [cite: shen2019nexus, hu2021scrooge, kim2023dream, crankshaw2017clipper, gujarati2020serving, yu2022orloj, ghafouri2023ipa]. Nexus [cite: shen2019nexus], Scrooge [cite: hu2021scrooge] and IPA [cite: ghafouri2023ipa] drop requests only when they cannot finish the current module inference within the SLO, overlooking latency budget requirements of subsequent modules. DREAM [cite: kim2023dream] simply assumes that the request will be executed with the smallest batch size without queueing and batch waiting in the subsequent module, leading to the drop too late issue. Additionally, Clipper [cite: crankshaw2017clipper], Orloj [cite: yu2022orloj] and Clockwork [cite: gujarati2020serving], are designed for single model tasks. These systems drop requests based on whether the request has already timed out or if the remaining latency is sufficient to complete the inference, and they can be adapted to serve inference pipelines via latency splitting. The dropping policies of all these systems share a reactive design, which utilizes only unidirectional information from previous modules and serves requests in order without considering their latency difference, leading to sub-optimal goodput.
}{%
    Numerous inference systems adopt dropping to ensure service quality~\cite{shen2019nexus, hu2021scrooge, kim2023dream, crankshaw2017clipper, gujarati2020serving, yu2022orloj, ghafouri2023ipa}. Nexus~\cite{shen2019nexus}, Scrooge~\cite{hu2021scrooge}, and IPA~\cite{ghafouri2023ipa} drop requests without considering downstream latency budgets, similar to \sysname-back, which wastes resources by executing requests already close to timeout. DREAM~\cite{kim2023dream} assumes requests will run with the smallest batch size and no queueing in later modules, similar to \sysname-sf, leading to drop-too-late issue. Orloj~\cite{yu2022orloj} and Clockwork~\cite{gujarati2020serving} target single-model workloads, dropping only when deadlines are missed or remaining budget is insufficient. Extending them to pipelines through latency splitting (\eg \sysname-split or \sysname-WCL) still yields reactive, stage-local designs that lack pipeline-wide awareness. 
    
}

\revise{
    Furthermore, several overload control techniques [cite: cho2020overload, zhou2018overload, welsh2002overload, welsh2001seda] for workloads such as microservices and DAGs can be regarded as a form of request dropping. These techniques reactively perform admission control based on queueing latency or system load. However, they cannot manage latency budgets across modules and handle latency uncertainty introduced by the batching process, hindering the ability to make appropriate dropping decisions for the inference pipeline.
}{%
    Beyond inference, several overload control techniques for microservices and DAG workloads~\cite{cho2020overload, zhou2018overload, welsh2002overload, welsh2001seda} can also be viewed as request dropping. These methods perform reactive admission control based on queueing delays or instantaneous load. However, like \sysname-oc, they neither handle latency budgets across modules nor address batching-induced uncertainty, limiting their effectiveness for inference pipelines.
    
}

\response{E}{Enhance related work with explicit baseline references.}

\label{page:related_latency}
\parab{\revise{}{Latency Profiling.}}
\revise{}{
    \sysname estimates per-request end-to-end latency by decoupling and profiling latency distributions to enable proactive dropping. Prior works~\cite{yu2022orloj, hao2017mittos, zhang2023shepherd, gujarati2020serving, cho2023protego} also profile request latency online but with different challenges and approaches. Orloj~\cite{yu2022orloj} captures empirical execution latency distributions in dynamic DNNs for request scheduling and batching to improve finish rates. Protego~\cite{cho2023protego} monitors contention queues to estimate per-request queueing delays and drop requests that would miss SLOs, thereby reducing tail latency. MittOS~\cite{hao2017mittos} predicts I/O request latency in the OS via queueing and device-efficiency profiling, rejecting requests unlikely to meet SLOs for early failover. While \sysname shares these insights on latency profiling and early rejection, it faces the unique challenge of highly uncertain batch wait times, amplified by model cascades. This motivates its pipeline-specific design, which balances over- and under-estimation to achieve high goodput.

    Other works~\cite{zhang2021sinan, gan2021sage, wang2024autothrottle} use profiled latency distributions to locate bottlenecks or manage resources rather than making per-request decisions. Nonetheless, their approaches inspire \sysname’s batch-wait estimation.
    

}

\response{B}{Expand discussion of related work on online latency distribution profiling.}

\parab{Request Priority.}
\revise{
    Several inference systems [cite: zhang2023shepherd, kannan2019grandslam] set different priorities for inference requests. For instance, Shepherd [cite: zhang2023shepherd] prioritizes requests with the closest deadlines, and GrandSLAm [cite: kannan2019grandslam] reorders requests in each module in descending order of the remaining latency budget. However, all these systems use a fixed request priority, limiting their ability to maintain high goodput across various workloads.
}{%
    Several inference systems~\cite{zhang2023shepherd, kannan2019grandslam} assign priorities to requests similar to \sysname-HBF and \sysname-LBF. For example, Shepherd~\cite{zhang2023shepherd} prioritizes requests with the closest deadlines, while GrandSLAm~\cite{kannan2019grandslam} reorders requests in descending order of remaining latency budget. However, these systems adopt fixed priority schemes, limiting their ability to sustain high goodput under diverse workloads.
    
}

\parab{Inference Optimizations.} Besides, numerous systems propose other inference optimization techniques for DNN serving, including resource scaling~\cite{cho2022sla, razavi2022fa2}, batching-aware scheduling~\cite{choi2021lazy, wu2023graft}, GPU scheduling~\cite{xia2023towards, pang2023efficient}, ensembling~\cite{bai2021automated, gunasekaran2022cocktail}, pipelining~\cite{jeong2023fast, kaler2022accelerating}, and spot instances~\cite{zhang2019mark, harlap2018tributary}. \sysname's proactive request dropping is orthogonal to these approaches and can be combined to further enhance the quality of service.

%% file: content/7_conclusion.tex
\section{Applicable to Other Workload}
\label{sec:discussion}

\begin{table}[t]
    \footnotesize
    \centering
    \setlength{\tabcolsep}{4pt}
    \caption{\revise{}{Evaluation setup of RAG workflow.}}
    \label{tab:rag_setup}
    \revise{}{
    \begin{tabular}{l l}
        \toprule
        \textbf{Setup} & \textbf{Description} \\
        \midrule
        \textbf{Testbed} & 2 $\times$ A100-80GB GPUs, vLLM v0.9.0~\cite{kwon2023efficient}, LangChain~\cite{langchain2025} \\
        \textbf{Input} & 10k queries from HotpotQA~\cite{yang2018hotpotqa} (Azure trace). \\
        \texttt{Rewrite} & Rewrite query with Llama-3-8B~\cite{grattafiori2024llama} (\textit{continuous batching}). \\
        \texttt{Retrieve} & \makecell[l]{Retrieve relevant context from FAISS~\cite{douze2024faiss} database, which\\contains 483k items from HotpotQA (\textit{batching execution})}\\
        \texttt{Search} & Search online with Tavily API~\cite{tavily2025} (\textit{multithreading}). \\
        \texttt{Generate} & Generate answer with Llama-3-8B (\textit{continuous batching}). \\
        \bottomrule
    \end{tabular}
    }
\end{table}

\begin{figure}[t]
    \centering
    \subfloat[Normalized Googput]{
        \includegraphics[width=0.46\linewidth]{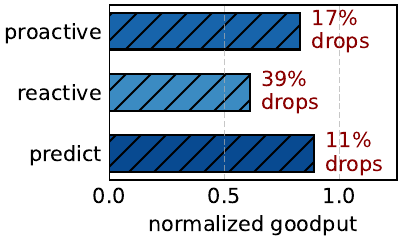}
        \label{fig:discussion_1}
    }\hspace{0pt}
    \subfloat[Module Latency Distribution]{
        \includegraphics[width=0.46\linewidth]{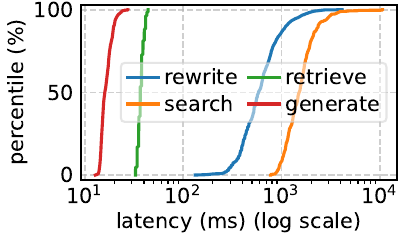}
        \label{fig:discussion_2}
    }
    \caption[]{\revise{}{Proactive drop demo in RAG workflow.}}\narrow{}
    \label{fig:discussion}
    \Description{}
\end{figure}

\label{page:discussion}
\revise{
    The core insight of \sysname, that proactively dropping certain requests can enhance overall goodput, demonstrates broad generalization potential beyond DNN inference pipeline workloads. Emerging domains such as Retrieval-Augmented Generation (RAG)~\cite{gao2023retrieval}, Large Language Model Agents~\cite{li2024survey}, and microservice workflows~\cite{luo2021characterizing} often exhibit similar multi-stage execution characteristics with strict latency SLOs. Thus, \sysname's proactive request dropping could offer a valuable paradigm for optimizing these diverse workloads. 
    
    However, effectively adapting \sysname's solution to these contexts necessitates addressing domain-specific challenges. For instance, latency estimation for autoregressive LLM generation and uncertain microservice execution path. Future investigations could adapt \sysname's core insight to these unique execution dynamics, potentially unlocking substantial goodput improvements in these emerging workloads.
}{%
    The core insight of \sysname, that proactively dropping certain requests can enhance overall goodput, generalizes beyond DNN inference pipelines. Emerging domains such as Retrieval-Augmented Generation (RAG)~\cite{gao2023retrieval}, Agentics~\cite{li2024survey}, and microservice workflows~\cite{luo2021characterizing} share multi-stage execution with strict latency SLOs. Thus, proactive dropping offers a promising paradigm for optimizing these workloads. 
    
    As a case study, we implement a four-module RAG pipeline as detailed in \tabref{tab:rag_setup}, where \texttt{retrieve} and \texttt{search} run in parallel as a DAG-style workflow. We set a time-to-first-token (TTFT) SLO of $5$s and compare two dropping policies: (1) \textit{proactive}, a customized version of \sysname's proactive dropping that estimates \texttt{rewrite} and \texttt{search} latencies by recent averages and estimates \texttt{generate}’s prefill latency using offline profiling and input length, while \texttt{retrieve} latency is estimated as in \sysname; and (2) \textit{reactive}, which drops requests only after exceeding the TTFT SLO. 
    
    As shown in \figref{fig:discussion_1}, proactive dropping reduces drop rate by $22\%$, confirming its applicability. However, \figref{fig:discussion_2} highlights key differences from DNN inference: \texttt{rewrite} latency varies with output length; \texttt{rewrite} and \texttt{generate} use continuous batching, eliminating batch wait; and \texttt{search} suffers long-tail latency from network delays. Consequently, even proactive dropping leaves $17\%$ of requests dropped. With oracle knowledge of \texttt{rewrite} output length (from offline runs with temperature $0$), the drop rate falls to $11\%$ (policy \textit{predict} in \figref{fig:discussion_1}). While unrealistic in practice, this can be approximated using output-length prediction strategies~\cite{qiu2024power, jin2023s3}. In summary, \sysname's proactive dropping generalizes to diverse pipeline workflows, but effective adaptation requires addressing domain-specific latency estimation challenges.

}

\response{B/D/E}{Extend the discussion with a case study on RAG pipeline.}

\section{Conclusion}
\label{sec:conclusion}

In this paper, we present \sysname, a DNN inference system designed to enhance goodput by proactively dropping certain requests to meet latency objectives under real-world workloads. \sysname leverages novel \revise{bi-directional}{proactive} request dropping and adaptive request priority methods to determine when to drop requests and which requests to drop at each module, optimizing goodput for the entire workload. Unlike existing systems' \textit{reactive} dropping policies, which suffer from drop-too-late and drop-wrong-set issues, \sysname's \textit{proactive} approach significantly improves the goodput. The evaluation shows that \revise{\sysname achieves $21\%$-$176\%$ higher goodput than the state of the art, while reducing the drop rate and invalid rate by $1.6\times$-$16.7\times$ and $1.5\times$-$62\times$ respectively.}{\sysname achieves $16\%$-$176\%$ higher goodput than the state of the art, while reducing the drop rate and invalid rate by $1.6\times$-$17\times$ and $1.5\times$-$62\times$ respectively.}

\section{Acknowledgments}

We appreciate the insightful feedback from the anonymous reviewers. This work is supported by the National Natural Science Foundation of China under grants No. 62572341.